\documentclass[a4paper,11pt]{article}
\pdfoutput=1
\usepackage{jheppub}
\usepackage{color,tensor}
\usepackage{graphicx}
\usepackage{wrapfig, enumerate, slashed}
\usepackage[noabbrev]{cleveref}
\usepackage{footmisc}
\usepackage{amsmath}
\usepackage{wasysym} 
\usepackage{graphicx}
\usepackage{color}
\usepackage{comment}
\usepackage{hyperref}
\usepackage{bbold}
\usepackage{comment}
\newcommand{\n}{\nonumber}

\newcommand{\bee}{\begin{eqnarray}}
\newcommand{\eee}{\end{eqnarray}}
\newcommand{\beeq}{\begin{equation}}
\newcommand{\eeeq}{\end{equation}}

\renewcommand{\vec}{\bf}
\newcommand{\iu}{{i\mkern1mu}}
\newcommand{\beqa}{\begin{eqnarray}}
\newcommand{\eeqa}{\end{eqnarray}}
\newcommand{\be}{\begin{equation}}
\newcommand{\ee}{\end{equation}}
\newcommand{\ba}{\begin{array}} 
\newcommand{\ea}{\end{array}}
\def\bea{\begin{eqnarray}}
\def\eea{\end{eqnarray}}
\newcommand{\nn}{\nonumber}

\newcommand{\Dfb}{\mathord{\buildrel{\lower3pt\hbox{$\scriptscriptstyle{\leftrightarrow \tiny{ \ \ \ } }$}}\over {D^{\mu}}}}
\newcommand{\Dfbd}{\mathord{\buildrel{\lower3pt\hbox{$\scriptscriptstyle\leftrightarrow$}}\over {D}_{\mu}}}

\title{
Gauge Choices, Infrared Pitfalls, and Thermal Effects in Effective Potentials
}
\author{Debanjan Balui$^{\dagger}$,}
\author{Tisa Biswas$^{\dagger}$,}
\author{Joydeep Chakrabortty$^{\dagger}$,}
\author{Debmalya Dey$^{\dagger}$,}
\author{Christoph Englert$^{\pounds}$,} 
\author{Subhendra Mohanty$^{\ddagger\dagger}$}
\affiliation{$^{\dagger}$Indian Institute of Technology Kanpur, Kalyanpur, Kanpur 208016, Uttar Pradesh, India}
\affiliation{$^{\pounds}$School of Physics \& Astronomy, University of Glasgow, Glasgow G12 8QQ, United Kingdom}
\affiliation{$^{\ddagger}$$^{\dagger}$IISER Bhopal, Bhopal 462066,  Madhya Pradesh, India}
\emailAdd{debanjanb22@iitk.ac.in}
\emailAdd{tisab@iitk.ac.in}
\emailAdd{joydeep@iitk.ac.in}
\emailAdd{debmalyad23@iitk.ac.in}
\emailAdd{christoph.englert@glasgow.ac.uk}
\emailAdd{subhendram@iiserb.ac.in}

\abstract{
The evaluation of effective potentials is critical for a range of phenomenological applications, including inflation, vacuum stability, and phase transitions. A drawback arises from the gauge-dependence of the effective potential. Furthermore, in theories with spontaneous symmetry breaking, the effective potential exhibits infrared (IR) divergences in the limit of vanishing Goldstone masses. By considering the multiplicative anomaly that arises due to non-factorisation of elliptic operators in the Fermi gauge when computing the effective potential at one-loop order, we demonstrate that its gauge independence and IR behaviour are improved to the corresponding findings of Landau gauge calculations simultaneously. The latter are straightforwardly and transparently reproduced using an approach that employs the Heat Kernel technique, thereby providing a shortcut to reflect anomaly-related cancellations from the outset. Our findings generalise to the treatment of the effective potential at finite temperature. In particular, the Heat Kernel extends gauge independence to any value of the expansion in mass over temperature.
}

\begin{document}
\maketitle
\allowdisplaybreaks
\flushbottom
\section{Introduction}
\label{sec:intro}
The effective potential for scalar fields~\cite{Coleman:1973jx} offers a consistent framework for analysing vacuum configurations, incorporating radiative corrections. It plays a central role in studies of cosmological phase transitions in the early universe. Importantly, such transitions may leave imprints in the form of stochastic gravitational waves, making the effective potential a quantity of experimental relevance. In the Standard Model (SM), the effective potential of the Higgs field receives contributions from heavy SM states and, possibly, beyond-SM (BSM) interactions~\cite{Sher:1988mj}. There is a long-standing drawback when considering this effective potential, first pointed out by Jackiw~\cite{Jackiw:1974cv}: Upon integrating gauge boson fluctuations, the effective potential can become gauge-dependent. This observation also applies to the calculation of the effective potential at finite temperature, as demonstrated by Dolan and Jackiw~\cite{Dolan:1974gu}. 

Given the gauge-dependence of the Higgs field itself, it is neither problematic nor surprising that the effective potential exhibits a gauge-dependence as well. A similar behaviour is observed more generally for Green's and vertex functions, whose gauge-dependence is typically not regarded as pathological. The transformation properties of these quantities under variations of the gauge fixing parameter are governed by an extended set of BRST symmetries~\cite{Kluberg-Stern:1974iel,Piguet:1984js,Kluberg-Stern:1974nmx,Nielsen:1975fs,DelCima:1999gg}, which lead to a set of relations known as Nielsen identities. These identities are helpful in many ways. For example, they provide an elegant avenue for identifying theoretically meaningful input data for a renormalisation programme (e.g.~through the definition of masses~\cite{Gambino:1999ai,Grassi:2001bz}). They can then be further used to formulate a gauge-{\emph{in}}dependent classification of `signal processes' when we seek to separate these from irreducibly (and coherently) contributing `backgrounds'~\cite{Goria:2011wa}.

When applied to the effective potential (the sum of all 1-PI vertex functions evaluated at vanishing external momenta, we denote this by $V$), the Nielsen identity famously reads~\cite{Nielsen:1975fs}
\begin{equation}
\label{eq:niels}
\left(\xi \frac{\partial}{\partial \xi} + C(\xi,\varphi) \frac{\partial}{\partial \varphi}\right) V(\varphi,\xi) = 0.
\end{equation}
Here, $\varphi$ denotes a scalar field and $\xi$ the gauge fixing parameter. It was pointed out by Nielsen~\cite{Nielsen:2014spa} that the gauge-dependence of $V$ can be exchanged for a general field redefinition $\varphi\to \Pi(\varphi,\xi,\xi_0)$, such that $\Pi$ obeys the differential equation Eq.~\ref{eq:niels} with boundary condition $\Pi(\varphi,\xi=\xi_0,\xi_0)=\varphi$ for a preferred gauge parameter $\xi_0$. Writing the potential in these new coordinates $V(\varphi,\xi)=V(\Pi(\varphi,\xi,\xi_0),\xi)$, Eq.~\ref{eq:niels} holds straightforwardly, and the potential no longer shows an explicit gauge-dependence by construction. The existence of such a constant $C$ could be considered a little ad hoc; it can be defined directly from the derivatives in Eq.~\ref{eq:niels}. However, the fact that this $C$ can alternatively (and more tediously) be derived from the generalised BRST transformation is nontrivial. 

Despite the gauge-dependence of the effective potential, it is still possible to extract some of its properties, such as its maxima and minima, as the gauge-dependence drops out at the extremal points of the potential~\cite{Aitchison:1983ns, Metaxas:1995ab}, cf. Eq.~\ref{eq:niels} for $\partial V/\partial \varphi=0$.  However, the gauge independence of these points can be subtle and relies on resummation as well as the renormalisation programme~\cite {Patel:2011th,Andreassen:2014eha}. Any non-linear field redefinition that removes gauge-dependencies from the effective potential will also be reflected in the gradient expansion beyond the effective potential term~\cite{Metaxas:1995ab,Alexander:2008hd,Garny:2012cg,Espinosa:2015qea}.

Applications to phase transitions require generalising the calculation of the effective potential to finite temperatures, which in the imaginary-time formalism involves taking the energy as a discrete set of Matsubara frequencies (for a review, see~\cite{Quiros:1999jp}). A sum over the Matsubara modes then replaces the integral over the energy of closed loops in Feynman diagrams. The application of the Nielsen identities at finite temperature to the calculation of phase transitions from effective potentials shows that while the finite temperature potential is, in general, gauge-dependent, again specific quantities like the critical temperature, which are defined at the extremal values of the potential, can be gauge-independent~\cite{Patel:2011th,Hirvonen:2021zej,Lofgren:2021ogg, Bernardo:2025vkz}. 

Furthermore, in theories with broken symmetries, the effective potential at the minima gives rise to massless Goldstone modes. In some gauges, such as the Fermi gauge, the effective potential diverges in the massless Goldstone limit. It was shown in~\cite{Garny:2012cg, Elias-Miro:2014pca, Espinosa:2016uaw} that after resumming the Goldstone mass term including loop corrections, the problem of gauge-dependence and infrared divergence is solved simultaneously, and the potential is gauge-independent at the minima in agreement with the Nielsen identity.

Recently, a novel approach~\cite{Balui:2025kat} has been proposed to resolve the issue of gauge-dependence of the effective potentials, where it was pointed out that the gauge-dependent term arises due to the incomplete factorisation of the regularised trace of a product of operators. The one-loop quantum corrections involve traces of elliptic operators, and these divergent traces are regularised using the zeta-function regularisation scheme~\cite{Hawking:1976ja}. A correct factorisation can necessitate the inclusion of the so-called multiplicative anomaly term~\cite{Elizalde2, Elizalde-3, Wodzicki, Mickelsson:1994fb, Bytsenko:1994bc, Elizalde:1997nd, Elizalde:1998xq}. It was shown in~\cite{Balui:2025kat} that the one-loop potential in the Fermi gauge, characterised by the gauge fixing term $(\partial^\mu A_\mu)^2/{2\xi} $, is susceptible to such a shortfall in factorisation, and including the multiplicative anomaly term precisely cancels the gauge-dependent term in the one-loop effective potential. This result is further corroborated by carrying out the one-loop integration of the heavy particles using the Heat Kernel method~\cite{Seeley,Belkov:1995gjw,Vassilevich:2003xt,Avramidi:2001ns,Kontsevich:1994xe,Avramidi:1990je,Banerjee:2023iiv,Chakrabortty:2023yke,Banerjee:2023xak,Banerjee:2024rbc,Adhikary:2025pbb,Adhikary:2025gdh}. Using the Heat Kernel method, it was also shown that upon integrating out gauge boson fluctuations, in the Fermi gauge, the gauge-dependent term in the one-loop effective potential is a total divergence consistent with the result obtained from the application of the multiplicative anomaly~\cite{Balui:2025kat}.

In this paper, we extend the results of~\cite{Balui:2025kat} and revisit the issue of gauge invariance in conjunction with infrared divergence in theories with symmetry breaking, taking into account multiplicative anomalies. We also compare this approach to computations based on the Heat Kernel method. We find that, starting with Abelian gauge theories coupled to scalars in the Fermi gauge, integrating out the massive gauge bosons and the Goldstone bosons, our result for the effective potential is independent of the gauge fixing parameter. This is equally true for non-Abelian gauge theories, as the self-interactions of gauge bosons in the non-Abelian theories have no contribution to the one-loop effective potential. We derive the effective potential for the Standard Model Higgs obtained by integrating out the massive fermionic and bosonic fluctuations starting with the $\xi$-dependent Fermi gauge and show that the one-loop effective potential is gauge invariant. This opens up applications of the effective potential to, e.g., problems of inflation and phase transitions~\cite{Urbano:2019ohp}, where the slope and rolling of the scalar over the entire potential are important and not just the extremal points. In parallel to gauge invariance, the singular infrared behaviour in the Fermi gauge is addressed by the multiplicative anomaly. Finally, we extend the derivation to the case of finite temperature. The finite temperature application is most naturally performed using the Heat Kernel method, which involves taking the trace of the zeroth component of momentum as a sum over discrete Matsubara frequencies. 
Our observations can therefore be understood as providing the theoretical justification for a practitioner's recipe to avoid gauge dependencies where they arise due to non-factorisation, making the application of Nielsen identities to remove gauge dependencies from the effective potential redundant. In parallel, infrared issues are avoided.

This paper is organised as follows. In Section~\ref{sec:SQEDEP}, we derive the effective potential for the Abelian Higgs theory in the Fermi gauge and show that the result is gauge invariant after considering the multiplicative anomaly. The Nielsen identity reduces to a statement of gauge invariance of the effective potential at all points, not just at the extrema of the potential. To keep our discussion transparent, the results for the full Standard Model are relegated to Appendix~\ref{sec:SMEP}. We demonstrate that, starting from the Fermi gauge, the one-loop effective potential for the Standard Model Higgs is gauge invariant. In Section~\ref{sec:HKEP}, we derive the one-loop potential using the Heat Kernel method. We extend our findings to the finite temperature case in Section~\ref{sec:CWFT}, and we summarise the results and give our conclusions in Section~\ref{sec:conclusion}.

\vspace{-0.18cm}
\section{Massive Scalar QED: Effective Potential}
\label{sec:SQEDEP}
We consider a massive scalar quantum electrodynamic theory (MSQED), in the Fermi gauge, expressed by the following Lagrangian~\cite{Jackiw:1974cv,Dolan:1974gu, Espinosa:2016uaw}
\begin{equation}\label{eq:SQEDLag}
  {\cal L}=-\frac14 F_{\mu\nu}F^{\mu\nu}-\frac{1}{2\xi}(\partial^\mu A_\mu)^2 + |D_\mu\Phi|^2 + m^2\Phi^\dagger\Phi - \lambda (\Phi^\dagger\Phi)^2 ,
\end{equation}
with $\Phi(x) \equiv \frac{1}{\sqrt{2}} (\phi_1 + i \phi_2)$.  
 The covariant derivative for MSQED is given by $D_\mu \equiv \partial_\mu - i g A_\mu$, with $A_\mu$ as the gauge field.
We shift the gauge field $A_\mu \to \eta_\mu$ and the scalar fields as $\phi_1 \to \varphi\; +\; h(x);\;\; \phi_2 \to  \chi(x)$. Here, $\varphi$ is a constant background around which $h(x)$ is the associated Abelian Higgs fluctuation. $\chi$ and $\eta_\mu$ are the Goldstone and gauge field fluctuations, respectively, around a vanishing background. The effective potential at one-loop, after integrating the scalar and gauge field fluctuations, is given as~\cite{Espinosa:2016uaw}
\begin{equation}
\label{eq:SQEDVeff}
  V_{\rm eff}^{(1)} (\varphi)  = i \;\sum_{n \, \in\{ \, \mbox{\tiny MSQED}\}} \; c_s  \int \frac{d^4k}{(2\pi)^4}  \log \det i{\tilde{\cal G}}^{-1}_n \{ \varphi; k \} , 
\end{equation}
where $c_s =-1/2$ for real bosonic fields. 
Here, we express the mass terms for the scalar and gauge fields in terms of the constant background field $\varphi$ as 
\begin{equation}\label{eq:SQEDmass}
   M_\chi^2 = -m^2 + \lambda \varphi^2 \,, \;
   M_h^2 = -m^2 + 3 \lambda \varphi^2 \,, \;
   M_A^2 = g^2 \varphi^2 \, .
\end{equation}
The Green's function $i\tilde{\mathcal{G}}^{-1}_n$ is determined from the quadratic fluctuation in the Lagrangian, and given as 
\begin{equation}
    \frac{1}{2} \Psi^{T} \left( i \mathcal{G}^{-1}_n \right) \Psi  = \frac{1}{2} 
    \begin{pmatrix}
        \eta_{\mu} & \chi & h 
    \end{pmatrix}
    \begin{pmatrix}
        i \tensor{\left( \mathcal{G}^{-1}_A \right)}{^\mu_\nu} & M_A \partial^{\mu} & 0 \\
        - M_A \partial_{\nu} & i \mathcal{G}^{-1}_{\chi} & 0 \\
        0 & 0 & i \mathcal{G}^{-1}_{h}
    \end{pmatrix}
    \begin{pmatrix}
        \eta^{\nu} \\
        \chi    \\
        h
    \end{pmatrix} \,,
\end{equation}
with $\Psi^T = \left( \eta_{\mu} , \chi ,  h \right) $.
The Green's function for the gauge boson is given in momentum space as
\begin{equation}
\label{defPropGauge}
    i \tensor{( \tilde{\mathcal{G}}^{-1}_A )}{^\mu_\nu}  =
    i \tilde{\mathcal{G}}^{-1}_{T} \,\left(\tensor{g}{^\mu_\nu} - \frac{k^\mu k_\nu}{k^2} \right)
    + i \tilde{\mathcal{G}}^{-1}_{L} \, \left( \frac{k^\mu k_\nu}{k^2} \right)
     ,
\end{equation}
with
\begin{equation}
\label{defPropGaugeL&T}
    i \tilde{\mathcal{G}}^{-1}_{T} = - k^2 + M_A^2 \,, \; 
    i \tilde{\mathcal{G}}^{-1}_{L} = - \xi^{-1} k^2 + M_A^2 \,.
\end{equation}
For the Abelian Higgs and Goldstone bosons, we obtain
\begin{equation}
\label{defPropHiggs&Goldstone}
    i \tilde{\mathcal{G}}^{-1}_{h} = - k^2 + M_{h}^2 \,, \; i \tilde{\mathcal{G}}^{-1}_{\chi} = - k^2 + M_{\chi}^2\,,
\end{equation}
respectively.

In dimensional regularisation $(d=4-2\epsilon)$, the one-loop effective potential (OLEP) reads \cite{Jackiw:1974cv,Dolan:1974gu, Espinosa:2016uaw}
\begin{eqnarray}
\label{EP1loopSMdrexplFermi}
    V^{(1)}_{\rm{eff}}|^{\rm{F}}_{\rm MSQED}  &=& -\frac{i}{2} \mu^{2\epsilon} \int \frac{d^d k}{(2\pi)^d} \Bigg[ 
    (d-1) \log \left( -k^2 + M_A^2 \right) + \log \left( k^2 - M_h^2 \right) \nonumber \\
    &+& \log \left( k^2 - M_{G_{+}}^2 \right) + \log \left( k^2 - M_{G_{-}}^2 \right) + \text{$\varphi$--independent terms} \Bigg]\,,
\end{eqnarray} 
where
\begin{equation} 
\label{defmassZFermi} 
{M}_{G_{\pm}}^2 = \frac{1}{2}  \left(  M_\chi^2 \pm \sqrt{ M_\chi^4 - 4 \xi M_\chi^2 M_A^2} \right) \,. 
\end{equation} 
Hereafter, we will consider only $\varphi$-dependent terms in the potential. Using the $\overline{\text{MS}}$ renormalisation scheme, the one-loop effective potential can finally be written as \cite{Jackiw:1974cv,Dolan:1974gu, Espinosa:2016uaw}
\begin{eqnarray}\label{eq:SQEDVeff1L} 
V^{(1)}_{\rm{eff}}|^{\rm{F}}_{\rm MSQED} &=& \frac{1}{64 \pi^2} \Bigg[
3 M_A^4 \left( \log \frac{M_A^2}{\mu^2} - \frac{5}{6} \right) 
+ M_h^4 \left( \log \frac{M_h^2}{\mu^2} - \frac{3}{2} \right) \nonumber \\
&&+ M_{G_+}^4 \left( \log \frac{M_{G_+}^2}{\mu^2} - \frac{3}{2} \right)   
+ M_{G_-}^4 \left( \log \frac{M_{G_-}^2}{\mu^2} - \frac{3}{2} \right) \Bigg]\,. 
\end{eqnarray}
The discussion presented here for the Abelian Higgs model can be extended to the more realistic case of the Standard Model along identical technical lines. The results are presented in App.~\ref{sec:SMEP}.

\subsection{Landau Gauge}
\label{subsec:LGSQED}
In the Landau gauge $\xi=0$, the one-loop effective potential, see Eq.~\ref{eq:SQEDVeff1L}, reads~\cite{Coleman:1973jx,Jackiw:1974cv,Dolan:1974gu,Fukuda:1975di, Elias-Miro:2014pca, Espinosa:2016uaw}
\begin{eqnarray}\label{eq:SQEDVG_LG}
V^{(1)}_{\rm eff}|^{\rm{L}}_{\rm MSQED}
&=& \frac{1}{64 \pi^2} \Bigg[
3 M_A^4 \left( \log \frac{M_A^2}{\mu^2} - \frac{5}{6}\right)
+ M_h^4 \left( \log \frac{M_h^2}{\mu^2} - \frac{3}{2}\right) \nonumber \\
&&
+ M_\chi^4 \left( \log \frac{M_\chi^2}{\mu^2} - \frac{3}{2}\right)
\Bigg]\,.
\end{eqnarray}
In this gauge and the Goldstone mass limit $M_\chi \to 0$ and $h \to 2\lambda \varphi^2$, the effective potential exhibits an infrared (IR) behaviour of the form $\simeq M_\chi^4 \, \log  (M_\chi^2/\mu^2)$. This expression is not problematic when this limit is taken, as both the potential and its first derivative are IR-safe, whereas higher-order derivatives of the potential are not. Using the resummation technique of~\cite{Elias-Miro:2014pca, Espinosa:2015qea, Espinosa:2016uaw}, any such IR divergences can be cured up to any order of perturbation. 
The Landau gauge results of the Abelian Higgs model are extended to the Standard Model case in App.~\ref{subsec:LGSM}.

\subsection{Fermi gauge}\label{subsec:FGSQED}
In the case of the Fermi gauge, $\xi$ is non-zero. To capture the dependence of $\xi$ on the effective potential, we consider the limit $4 \xi M_A^2 \gg M_\chi^2$. This assumption is consistent with the vanishing Goldstone mass limit, which reads $M_{G_\pm}^2 \simeq \frac{M_\chi^2}{2} \pm i \sqrt{\xi M_\chi^2 M_A^2}$. 
The Goldstone mass-dependent term in the effective potential in the Fermi gauge is given as
\begin{eqnarray}\label{eq:SQEDVGFG}
 V^{(1)}_{\rm eff} [M_\chi^2]
&=& \frac{1}{64 \pi^2} \left[M_{G_+}^4 \left( \log \frac{M_{G_+}^2}{\mu^2} - \frac{3}{2} \right) + M_{G_-}^4 \left( \log \frac{M_{G_-}^2}{\mu^2} - \frac{3}{2} \right) \right] \nonumber \\
&\simeq& \frac{1}{64 \pi^2}\left[ 2 \left( \frac{M_\chi^4}{4} - \xi M_\chi^2 M_A^2\right) \left(  \log \frac{|M_{G_+}^2|}{\mu^2} -\frac{3}{2} \right) \right]\,.
 \end{eqnarray}
The effective potential at one-loop order in the Fermi gauge reads~\cite{Andreassen:2014eha,Andreassen:2014gha,Espinosa:2015qea, Espinosa:2016uaw,Espinosa:2016nld}
\begin{eqnarray}\label{eq:SQEDVFG_1}
\tilde{V}^{(1)}_{\rm eff}|^{\rm{F}}_{\rm MSQED}
&=& \frac{1}{64 \pi^2} \Bigg[ 3 M_A^4 \left( \log  \frac{M_A^2}{\mu^2}  - \frac{5}{6}\right) 
+ M_h^4 \left( \log  \frac{M_h^2}{\mu^2}  - \frac{3}{2} \right) \nonumber \\
&& + 2\left( \frac{M_\chi^4}{4} - \xi M_\chi^2 M_A^2 \right)
      \left( \log \frac{|M_{G_+}^2|}{\mu^2}
      -\frac{3}{2} \right) \Bigg]\,.
\end{eqnarray}
In the Fermi gauge and the above limit of the vanishing Goldstone mass $M_\chi \to 0$ and $h\to 2\lambda \varphi^2$, the IR behaviour is modified to $\simeq M_\chi^2 \, \log  (M_\chi^2/\mu^2)$. This is more troubling than the Landau gauge result. Although the limit can be safely approached for the potential itself, all of its derivatives are IR-divergent. To improve this, one can employ the Nielsen identity to recover IR behaviour as $\simeq M_\chi^4 \, \log  \frac{M_\chi^2}{\mu^2}$ as encountered in the Landau gauge~\cite{Martin:2014bca, Espinosa:2015qea, Espinosa:2016uaw}; resummation ensures that the effective potential remains IR-safe up to all orders.  
The Fermi gauge results of the Abelian Higgs model are extended to the Standard Model case in App.~\ref{subsec:FGSM}

\subsection{Fermi Gauge and the Multiplicative Anomaly}
\label{subsec:FGMASQED}
We note that the computation of the one-loop effective potential involved factorising the operators for the Green's functions of gauge, Higgs, and Goldstone bosons. It has been pointed out that the simple factorisation result of unregularised operators within functional determinant $\text{Det}[\Delta_1 \; \Delta_2]= \text{Det}[\Delta_1] \,\text{Det}[ \Delta_2]$ is not valid, related to the noncommutative nature of the $\zeta$-trace. The $\zeta$-function regularisation \footnote{It is not restricted to $\zeta$-regularisation, and also appears in the case of dimensional regularisation \cite{Barvinsky:2024irk}.} introduces an additional factor called the `multiplicative anomaly'~\cite{Elizalde:1997nd}.

Specifically, let us consider the non-factorisation of the elliptic operators $(\Delta_i)$ in the functional method, and the requirement of including the `multiplicative anomaly'
$\mathbb{A}[\Delta_1, \Delta_2]$ as follows~\cite{Elizalde:1997nd,Kontsevich:1994xe, Bytsenko:1994bc,Bytsenko:2003}
\begin{eqnarray}\label{eq:MA-1}
\text{log[Det}[\Delta_1 \; \Delta_2]] =  \text{log[Det}[\Delta_1]]+\text{log[Det}[\Delta_2]] + \mathbb{A}[\Delta_1, \Delta_2]\,. 
\end{eqnarray}
We consider two elliptic operators of the form: $\Delta_i \equiv (k^2+\mathcal{C}_i^2);\;\; i=1,2$, then the anomaly is given for $d-$dimensional Euclidean space as \cite{Kontsevich:1994xe, Bytsenko:1994bc,Wodzicki,Mickelsson:1994fb,Bytsenko:1994bc,Elizalde:1997nd,Bytsenko:2003, Balui:2025kat}  
\begin{eqnarray}\label{eq:MA-2}
	\mathbb{A}[\Delta_1, \Delta_2] \supset \frac{(-1)^q\; \mathcal{V}_d}{4 (4\pi)^q \Gamma(q)} \sum_{n=1}^{q-1} \frac{1}{n(q-n)} [(\mathcal{C}_1^2)^n-(\mathcal{C}_2^2)^n] [(\mathcal{C}_1^2)^{q-n}-(\mathcal{C}_2^2)^{q-n}]\,,
\end{eqnarray}
with $q=d/2>1$, and $\mathcal{V}_d$ is the Euclidean volume.

This should be added to the effective potential computed in the Fermi gauge, and we will see that it serves the purpose of restoring gauge invariance and IR safety at the considered order. To this end, we focus primarily on extracting the $\xi$-dependent anomaly density term in $d=4$ Euclidean space~\cite{Bytsenko:2003, Balui:2025kat} that reads 
\begin{eqnarray}\label{eq:SQEDMA}
\mathbb{a}[M_{G_+}^2, M_{G_-}^2] (\xi) &=&		\frac{\mathbb{A}[M_{G_+}^2, M_{G_-}^2] (\xi)}{\mathcal{V}_4} \nonumber \\
&\supset& \frac{(-1)^q}{4 (4\pi)^q \Gamma(q)} \sum_{n=1}^{q-1} \frac{1}{n(q-n)} [(M_{G_+}^2)^n-(M_{G_-}^2)^n] [(M_{G_+}^2)^{q-n}-(M_{G_-}^2)^{q-n}] \nonumber\\ 
&=& \frac{1}{64 \pi^2}  \left(-4 \xi M_\chi^2 M_A^2 \right)\,.
\end{eqnarray}
%
The one-loop effective potential after incorporating the multiplicative anomaly\footnote{The factor ${1}/{2}$, in front of the anomaly density $\mathbb{a}$ is the same $c_s$ factor that appeared in Eq.~\ref{eq:SQEDVeff}.}~\cite{Balui:2025kat}, therefore, becomes
\begin{eqnarray}   \label{eq:SQEDVefMA}
	V_{\rm eff}^{(1)}|^{\rm{F+MA}}_{\rm MSQED} &=& \tilde{V}^{(1)}_{\rm eff}|^{\rm{F}}_{\rm MSQED}	- \frac{1}{2} \; \mathbb{a}[M_{G_+}^2, M_{G_-}^2] (\xi)~  \left(  \log  \frac{|M_{G_+}^2|}{\mu^2}  - \frac{3}{2} \right)\nonumber \\
    &=&\frac{1}{64 \pi^2} \Bigg[ 3 M_A^4 \left( \log  \frac{M_A^2}{\mu^2}  - \frac{5}{6}\right) 
+ M_h^4 \left( \log  \frac{M_h^2}{\mu^2}  - \frac{3}{2} \right) \nonumber \\
&& +~2\left( \frac{M_\chi^4}{4} - \xi M_\chi^2 M_A^2 \right)
      \left( \log  \frac{|M_{G_+}^2|}{\mu^2}
      -\frac{3}{2} \right) \Bigg] \nonumber \\
    &&- \frac{1}{2} \times \frac{1}{64 \pi^2} \left[ (- 4 \xi  M_\chi^2 M_A^2) \left( \log  \frac{|M_{G_+}^2|}{\mu^2}  - \frac{3}{2} \right)\right]  \nonumber \\
    &\simeq &\frac{1}{64 \pi^2} \bigg[ 3 M_A^4 \left( \log  \frac{M_A^2}{\mu^2}  - \frac{5}{6}\right)+ M_h^4 \left( \log  \frac{M_h^2}{\mu^2}  - \frac{3}{2} \right) \nonumber \\
    &&+ \frac{M_\chi^4}{2}  \left( \frac{1}{2} \log \frac{M_\chi^2}{\mu^2}  - \frac{3}{2} \right) \bigg]\,, 
\end{eqnarray}
where we discard the terms that are not relevant for our analysis.
After including the multiplicative anomaly, the IR behaviour in this effective potential now reproduces the IR behaviour found in the Landau gauge $\simeq M_\chi^4 \, \log  (M_\chi^2/\mu^2)$ with the consequences for the potential's derivatives as discussed above. 
Again, the results presented here for the Abelian Higgs model can be extended to the SM, which is detailed in App.~\ref{subsec:FGMASM}.

\subsection{Nielsen Identity with Multiplicative Anomaly}
\label{subsec:NIMASQED}
As discussed in the previous sections, neglecting the multiplicative anomaly leads to a gauge-dependent effective potential~\cite{Jackiw:1974cv}. Nielsen identity, as given in Eq.~\ref{eq:niels}, asserts that the information derived {\emph{only}} from the extremal points of the potential is gauge-independent. In this section, we, therefore, explore the connection between the multiplicative anomaly and the Nielsen identity.

The Nielsen identity in the given order in the loop-parameter $\hbar$ is given by \cite{Nielsen:1975fs}
\begin{equation}\label{eq:SQEDNI}
\xi \frac{\partial V^{(1)}}{\partial \xi} + C^{(1)} \frac{\partial V^{(0)}}{\partial \varphi} =0
\end{equation}
The coefficient $C^{(1)}$ is the one-loop contribution to a gauge transformation and has been calculated for the Abelian-Higgs model \cite{Aitchison:1983ns}. In this section, we compute this coefficient directly from the one-loop potential and show that there are two contributions; the first term is the same as in \cite{Aitchison:1983ns}, but there is an equal contribution from the multiplicative anomaly, which cancels this term and the net $C^{(1)}=0$. Therefore, the Nielsen identity, while formally correct, is reduced simply to the statement of gauge invariance of the one-loop effective potential to all orders of coupling.

The relevant Goldstone mass-dependent term of the effective potential, which contains the $\xi$ parameter, is given by
\begin{equation}    \label{eq:SQEDVG}
    \tilde{V}^{(1)}_{\rm eff}|^{\rm F}_{\rm MSQED} (|M_{G_+}^2|) \equiv -\frac{i}{2} \int \frac{d^4 k}{(2 \pi)^4} \log \left[ (k^2-M_{G_+}^2) (k^2 - M_{G_-}^2) \right] \,,
\end{equation}
so that
\begin{eqnarray}
\xi \frac{\partial \tilde{V}_{\rm eff}^{(1)}|^{\rm F}_{\rm MSQED}}{\partial \xi} &=& \frac{i}{2} \int  \frac{d^4 k}{(2 \pi)^4}\left( \xi \frac{\partial M_{G_+}^2}{\partial \xi} \frac{1}{(k^2-M_{G_+}^2)}+\xi \frac{\partial M_{G_-}^2}{\partial \xi} \frac{1}{(k^2-M_{G_-}^2)}\right) \n \\
&\simeq& -\frac{1}{64 \pi^2} (2 \xi M_\chi^2 M_A^2) \left( \frac{1}{2} \log \frac{M_\chi^2}{\mu^2} \right) \,.
\label{dvxi-1}
\end{eqnarray}
From this point on, we focus on the terms $\sim M_\chi^2 \log {M_\chi^2}/{\mu^2}$, which are relevant for the effective potential.
In addition, there is a $\xi$-dependent part in the multiplicative anomaly term (Eq.~\ref{eq:SQEDMA})
\begin{equation}\label{MAxi}
    \mathbb{a}[M_{G_+}^2, M_{G_-}^2] (\xi)=\frac{1}{64 \pi^2}  \left(-4 \xi M_\chi^2 M_A^2 \right) \,.
\end{equation} 
From Eqs.~\ref{dvxi-1} and \ref{MAxi} we see that the total $\xi$
dependence of the effective potential is 
\be\label{eq:dvxitot}
\xi \frac{\partial V^{(1)}}{\partial \xi}=\xi \frac{\partial \tilde{V}_{\rm eff}^{(1)}|^{\rm F}_{\rm MSQED}}{\partial \xi} - \xi\frac{d}{d\xi} \bigg[ \frac{1}{2} \times \mathbb{a}[M_{G_+}^2, M_{G_-}^2] (\xi) \left( \frac{1}{2} \log \frac{M_\chi^2}{\mu^2} \right) \bigg]=0
\ee
Comparing Eq.~\ref{eq:dvxitot} with the general Nielsen identity Eq.~\ref{eq:SQEDNI}, we see that for the Abelian Higgs model, the Nielsen coefficient is 
\be
 C^{(1)} \frac{\partial V^{(0)}}{\partial \varphi}=-\xi \frac{\partial V^{(1)}}{\partial \xi}=0.
\ee
In the standard calculations, for example in~\cite{Aitchison:1983ns}, the anomaly contribution is not taken into account. The expression for $C^{(1)}\partial V^{(0)}/\partial \varphi$ matches our expression in Eq.~\ref{dvxi-1}, taking the appropriate limits for the Abelian model in the Fermi gauge.
Conversely, we see that, in Eq.~\ref{eq:dvxitot}, the anomaly contribution to the $\xi$-dependent term cancels the $\xi$-dependent term in the standard one-loop calculation, and the total effective potential is gauge invariant at all field values and not just at the extrema of the potential. The Nielsen identity reduces to the statement that the one-loop effective potential is gauge invariant. This also demonstrates that the consideration of the multiplicative anomaly addresses gauge-dependence and IR sensitivity of the effective potential simultaneously in the Fermi~gauge. 

It is important to emphasise that the multiplicative anomaly stems from addressing the non-factorisation of elliptic operator traces, whereas removing gauge-dependence through redundant field redefinitions is an attempt to eliminate an issue that, as shown, does not arise in the first place.
The statements of this section generalise again to the Standard Model, which is detailed in App.~\ref{subsec:NIMASM}.

\section{Effective Potential via the Heat Kernel}
\label{sec:HKEP}
We will now compare our previous analysis with an investigation of the effective potential using the Heat Kernel (HK) method. HK~\cite{Seeley:1967ea, Belkov:1995gjw, Vassilevich:2003xt, Avramidi:2001ns, Kontsevich:1994xe, Banerjee:2023iiv, Banerjee:2023xak, Chakrabortty:2023yke} is defined as the solution of the heat equation
\begin{equation}\label{eq:heat_eq}
    \left(\partial_t+\Delta_x\right)K(t,x,y,\Delta)=0\,,
\end{equation}
with initial condition $ K(0,x,y,\Delta)=\delta(x-y)$. Here, 
\begin{equation}
\Delta_{ij} \equiv\frac{\delta^2 \mathcal{L}}{\delta \Phi_i \delta \Phi_j} = (D^2+M^2)_{ij} + U_{ij} 
\end{equation} 
is the elliptic operator in association with the scalar field $\Phi$. $M^2$ is diagonal non-degenerate matrix where $U$ captures all interactions. 
Within this framework, the HK is given as~\cite{Banerjee:2023xak, Chakrabortty:2023yke}
\begin{eqnarray}\label{eq:ZHK}
    \text{tr}\,K(t,x,x,\Delta) \! &=& \! \text{tr} \int \frac{d^4p}{(2 \pi)^4} \langle x| e^{-M^2 t}\, \mathcal{T} \text{exp}\Big[ -  \int_0^t \big(D^2 + e^{M^2 t'} U e^{-M^2 t'} \big) dt' \Big] |p \rangle \langle p|x \rangle \nonumber \\
    \! &=& \! \text{tr} \int \frac{d^4p}{(2 \pi)^4} e^{-M^2 t} \, e^{p^2 t} \mathcal{T} \text{exp}\Big[ \! - \! \int_0^t \big(D^2 + 2ip\cdot D + e^{M^2 t'} U e^{-M^2 t'} \big) dt' \Big]\, 
\end{eqnarray}
where we have used the property $e^{-i p \hat x} D_\mu  e^{i p \hat x}= D_\mu+ip_\mu$. Here, the trace is defined over all internal symmetry indices and the number of fields. After redefining $p^2 t \to p^2$, we find~\cite{Banerjee:2023xak, Chakrabortty:2023yke}
\begin{eqnarray}
    \text{tr} \, K(t,x,x,\Delta) = \text{tr} \int \! \frac{d^4p}{(2 \pi)^4\, t^2} e^{-M^2 t} e^{p^2 t} \, \mathcal{T} \text{exp}\Big[ \! - \! \int_0^t \big(D^2 + 2ip\cdot D/\sqrt{t} + e^{M^2 t'} U e^{-M^2 t'} \big) dt' \Big]\,. 
\end{eqnarray}
Here, $p^2= \delta_{\mu \nu}p^\mu p^\nu = -(p_1^2 + p_2^2 + p_3^2 + p_4^2) $.
To perform this integral perturbatively, we define the Schwinger-time ($t$) ordered $(\mathcal{T})$ exponential as 
\begin{eqnarray}
    \mathcal{F}(t,\mathcal{A}) &=& \mathcal{T} \text{exp}\big(-\int_0^t \mathcal{A}(t') dt' \, \big)
    =1 + \sum_{n=1}^{\infty} (-1)^n f_n(t,\mathcal{A})\,, 
\end{eqnarray}
where $f_n$ is expressed as the $t$-ordered Volterra integral equation of the second kind as
\begin{eqnarray}
    f_n(t,\mathcal{A})=\int_0^t ds_1 \int_0^{s_1} ds_2\, ...\int_0^{s_{n-1}} ds_n \mathcal{A}(s_1)\mathcal{A}(s_2)\,... \, \mathcal{A}(s_n) \,.
\end{eqnarray}
In case of two non-degenerate fields $(\Phi_1, \Phi_2)$, we define the $\mathcal{A}$-matrix as
\begin{eqnarray}
    \mathcal{A}(t') =\! \begin{pmatrix}
       D^2 + 2ip.D/\sqrt{t} + U_{11} \! & \! U_{12} \, e^{\Delta_{12}^2 t'} \\
       U_{21} \, e^{\Delta_{21}^2 t'} \!  & \! D^2 + 2ip.D/\sqrt{t} + U_{22}
   \end{pmatrix},
\end{eqnarray}
where $\Delta_{12}^2 = M_1^2 - M_2^2 = - \Delta_{21}^2$.

After performing the Schwinger-time integral, we can compute the one-loop effective Lagrangian as follows~\cite{Banerjee:2023xak, Chakrabortty:2023yke}
\begin{eqnarray}\label{eq:HKLeff}
    \mathcal{L}_{eff} &=& c_s\; \text{tr}\, \int_0^{\infty} \frac{dt}{t} K(x,x,\Delta,t) 
    = c_s \; \text{tr}\, \int_0^\infty \frac{dt}{t^3} \int \frac{d^4p}{(2 \pi)^4} e^{-M^2 t} e^{p^2 t} \big[ 1 + \sum_n (-1)^n f_n(t,\mathcal{A})  \big]\,,
\end{eqnarray}
where $c_s=1/2$ or $1$ for real or complex bosons, respectively.

In the following subsection, we will compute the one-loop effective potential for the previously discussed scenario: MSQED (see Sec.~\ref{sec:SQEDEP}), employing the HK method. It is important to note that, as we will deal with the full matrix elliptic operator, unlike the usual functional methods, there is no source of multiplicative anomaly. 
We discuss the same for the SM in Sec.~\ref{sec:SMEP}.
\subsection*{Massive Scalar QED}
\label{sec:HKSQED}
We start with the same MSQED Lagrangian, Eq.~\ref{eq:SQEDLag}, which reads as
\begin{equation}
  {\cal L}=-\frac14 F_{\mu\nu}F^{\mu\nu}-\frac{1}{2\xi}(\partial^\mu A_\mu)^2 + |D_\mu\Phi|^2 + m^2\Phi^\dagger \Phi - \lambda (\Phi^\dagger \Phi)^2  \,.
\end{equation}
Here, we define the elliptic operator in the basis $\Psi=(\eta_\mu,\chi,h)^T$, defined in Sec.~\ref{sec:SQEDEP}, as
\begin{equation}
    \Delta_{ij}=\dfrac{ \delta^2 \mathcal{L}}{\delta \Psi_i \delta \Psi_j} =  (\partial^2 + M^2)_{ij} + U_{ij} \,,
\end{equation} 
where
    \begin{equation}
     M^2 =
\begin{pmatrix}
 g^2 \varphi^2 \delta_{\mu \nu} & 0 & 0 \\
0 & -m^2 + \lambda \varphi^2 & 0 \\
0 & 0 & -m^2 + 3 \lambda \varphi^2
\end{pmatrix}\,,
\end{equation}
and 
	\begin{equation} U  = \begin{pmatrix}
 \left(1- \frac{1}{\xi}\right)\partial_\mu\partial_\nu & \, g\partial_\mu \varphi & \, 0 \\[6pt]
- g \partial_\nu \varphi & \, 0 & \,0 \\[6pt]
0 & 0 & \, 0
\end{pmatrix}\,.
\end{equation}  
Here, we have the $\mathcal{A}$-matrix in the following form
\begin{multline} \label{eq:SQED-A}
    \mathcal{A} =\\ 
    \resizebox{0.97\textwidth}{!}{$
    \begin{pmatrix}
      \, (\partial^2 \!+\! 2ip\cdot\partial/\sqrt{t} ) \delta_{\mu \nu} \! + \! (1\!-\! \frac{1}{\xi})(\partial_\mu \partial_\nu - p_\mu p_\nu/t + 2ip_\mu \partial_\nu/\sqrt{t})  & \; \; g \partial_\mu \varphi + ig p_\mu \varphi/\sqrt{t} \, &  0 \\
      \, -g \partial_\nu \varphi - ig p_\nu \varphi/\sqrt{t} &  \partial^2 \!+\! 2ip\cdot\partial/\sqrt{t}\! & 0\\
        0  &  0 & \partial^2 \!+\! 2ip\cdot\partial/\sqrt{t}\!
   \end{pmatrix}$}\,.
\end{multline}
After incorporating $\mathcal{A}$, Eq.~\ref{eq:SQED-A}, in effective Lagrangian, Eq.~\ref{eq:HKLeff}, we find the one-loop effective potential as
\begin{eqnarray}\label{eq:Veff-SQED-HK}
	V_{\rm eff}^{(1)}|^{\rm{HK}}_{\rm MSQED} &=&  \frac{1}{ 64 \pi^2} \Big[ M_h^4\; \Big( \text{log} \,  \frac{M_h^2}{\mu^2}  -\frac{3}{2} \Big) + M_\chi^4 \; \Big( \text{log}\,  \frac{M_\chi^2}{\mu^2} -\frac{3}{2} \Big) + 3\;M_A^4 \;\Big( \text{log} \, \frac{M_A^2}{\mu^2} -\frac{5}{6} \Big) 
    \nonumber \\
&& + \left(1 - \frac{1}{\xi} \right)^2 \,  \left( \log  \frac{M_A^2}{\mu^2} \right)\; \partial^4
+ 2 \left( 1 - \frac{1}{\xi} \right) \, \left( \log  \frac{M_A^2}{\mu^2}  - 1 \right) \; \partial^2 M_A^2 \Big]\,,
\end{eqnarray}
where field-dependent mass terms are
\begin{equation}
    M_h^2 = - m^2 + 3 \lambda \varphi^2 \,\, , \, M_\chi^2 = -m^2 + \lambda \varphi^2 \,\, , \, M_A^2 = g^2 \varphi^2 \,.
\end{equation}

It is important to note that the gauge parameter $\xi$ is associated with two terms in Eq.~\ref{eq:Veff-SQED-HK}: 
$\partial^4 $, and $\partial^2 M_A^2$, which are total derivatives and vanish identically for constant background fields~\cite{Balui:2025kat}. Thus, the relevant effective potential reads 
\begin{eqnarray}\label{eq:Veff-SQED-HK-final}
	V_{\rm eff}^{(1)}|^{\rm{HK}}_{\rm MSQED} &=&  \frac{1}{ 64 \pi^2} \Big[ M_h^4 \; \Big( \text{log} \,  \frac{M_h^2}{\mu^2}  -\frac{3}{2} \Big) + M_\chi^4 \; \Big( \text{log}\, \frac{M_\chi^2}{\mu^2} - \frac{3}{2} \Big) + 3\;M_A^4 \; \Big( \text{log} \, \frac{M_A^2}{\mu^2} -\frac{5}{6} \Big) 
   \Big]\,, 
\end{eqnarray}
which exactly matches the same computation employing the functional method in the Landau gauge, see Eq.~\ref{eq:SQEDVG_LG}. We apply the HK method to the SM in App.~\ref{sec:HKSM}.

\section{Effective Potential at Finite Temperature}
\label{sec:CWFT}
The implications of the previous sections can be generalised to finite temperature. In this section, we compute the one-loop Coleman-Weinberg (CW) effective potential at finite temperature for a massless scalar QED (SQED)~\cite{Dolan:1973qd,Jackiw:1974cv}.
In the earlier sections, we have established that the result for the effective potential obtained using the functional method, augmented with a multiplicative anomaly, is the same as when the computation is performed using the Heat Kernel method. The effective potential at zero temperature in the Landau gauge ($\xi=0$) is the same as the result of the Fermi gauge calculation in the Heat Kernel method. In this section, we use the finite temperature extension of the Heat Kernel method \cite{Megias:2003ui,Chakrabortty:2024wto}. We observe that the finite temperature effective potential in the $\xi=0$ gauge yields the same result as an equivalent calculation performed using the Heat Kernel method, starting from the Fermi gauge. 

The SQED Lagrangian in Fermi gauge, at zero temperature, defined in $R^3 \times R^1$\footnote{We will work with $d+1=4$ dimensional Euclidean manifold.} space-time manifold reads as (see Eq.~\ref{eq:SQEDLag})
\begin{equation}\label{eq:SQEDL}
  {\cal L}=-\frac14 F_{\mu\nu}F^{\mu\nu}-\frac{1}{2\xi}(\partial^\mu A_\mu)^2 + |D_\mu\Phi|^2  - \lambda (\Phi^\dagger\Phi)^2\, ,
\end{equation}
with $\Phi(x) \equiv \frac{1}{\sqrt{2}} (\phi_1 + i \phi_2)$. Here, we consider $\phi_i = \hat{\phi} + \eta_i$ and $A_\mu = \eta_\mu$ where $\eta_i$ are the scalar fluctuations around a constant background $\hat{\phi}$, and $\eta_\mu$ is the gauge fluctuation around zero background. To compute the effective action at finite temperature $(T)$, we will use the imaginary time formalism where the space-time manifold becomes $R^3 \times S^1$ with $\beta$ as the radius of $S^1$. The temperature is then defined as the inverse radius of $S^1$, i.e., $ T = 1/\beta$. The quantum fields satisfy the following periodic boundary condition along the temporal direction, i.e., on $S^1$
\begin{equation}
    \Psi (t+\beta, {\vec x}) = \pm    \Psi (t, {\vec x})\,,
\end{equation}
where $+$ and $-$ are for bosonic and fermionic fields, respectively. Here, we work with the temporal gauge $A_0=0$.\footnote{Though temporal gauge is not the best choice~\cite{Megias:2003ui, Moral-Gamez:2011wcb, Chakrabortty:2024wto}, it is simpler to implement, and our conclusion in this work remains unaffected by this choice.}

\subsection{Effective Potential with the Functional Method: Finite Temperature}
\label{subsec:FT-FM}
Following the prescription of Jackiw and Dolan~\cite{Dolan:1973qd}, after integrating out the quantum fluctuations at finite temperature, we find the one-loop effective potential in Fermi gauge as
\begin{eqnarray}\label{V_eff_temp}
    V_{\text{CW}}^{\beta}|^{{\rm FG}} (\hat \phi) &=& -\frac{i}{2}  \int\frac{d^4k}{(2\pi)^{4}} \,\text{log}\, {\rm det} \, i \mathcal{D}^{-1}\{\hat{ \phi}; k\} \\ \nonumber
    &=& \frac{1}{2 \beta} \sum_{n=-\infty}^{\infty}\int\frac{d^3k}{(2\pi)^{3}} \,\text{log}\,[(\omega_n^2+{\vec k}^{2} + \mathcal{A}_{1}^{2})(\omega_n^2+{\vec k}^{2} + \mathcal{A}_{2}^{2})(\omega_n^2+{\vec k}^{2} + \mathcal{A}_{3}^{2})(\omega_n^2+{\vec k}^{2} + \mathcal{A}_{4}^{2})^3]\,,
\end{eqnarray}
where the Matsubara frequency $\omega_n = {2\beta^{-1} \pi n }$, and 
\begin{eqnarray}
	\mathcal{A}_{1}^{2} &=& \frac{1}{12}(\lambda\hat{\phi}^{2} + \hat{\phi}^{2}\sqrt{\lambda^{2} - 24\xi\lambda e^{2}})\,,\quad
	\mathcal{A}_{3}^{2} = \frac{1}{2}\lambda\hat{\phi}^{2}\,,\quad  \\
	\mathcal{A}_{2}^{2} &=& \frac{1}{12}(\lambda\hat{\phi}^{2} - \hat{\phi}^{2}\sqrt{\lambda^{2} - 24\xi\lambda e^{2}})\,,\quad 
\mathcal{A}_{4}^{2} = e^{2}\hat{\phi}^{2}\,.
\end{eqnarray}
Here,
\begin{equation}\label{eq:GreenFunction}
    \mathcal{D}_{pq} = \frac{\delta^2 \mathcal{L}}{\delta \Psi_p \delta \Psi_q} = 
    \begin{pmatrix}
        (\partial^2 + \frac{\lambda}{6} \hat{\phi}^2 ) \delta_{ab} + \frac{\lambda}{3} \hat{\phi}_a \hat{\phi}_b & e \epsilon_{ab^\prime} \partial_\nu\hat{\phi}_{b^\prime} \\
        -e \epsilon_{a^{\prime}b} \partial_\mu \hat{\phi}_{a^{\prime}} & -(\partial^2 + e^2 \hat{\phi}^2 ) \delta_{\mu \nu}  + (1- \frac{1}{\xi}) \partial_\mu \partial_\nu
    \end{pmatrix}\,,
\end{equation}
with $\Psi_p \in \{\phi_a,A_\mu \}$, $a=1,2$, and $\mu=0,1,2,3$.
The finite part of the effective potential in the $d+1=4-2\epsilon$ dimension is given as 
\begin{equation}
 V_{\text{CW}}^{\beta}|^{{\rm FG}} (\hat{ \phi}) = - \frac{1}{2 \beta} \frac{\mu^{2 \epsilon} \; \Gamma \left( -\frac{d}{2} \right)}{\left( 4 \pi \right)^{\frac{d}{2}}} \sum_{n} \Bigg[ \left( \omega_{n}^2 \; + \; \mathcal{A}_1^2 \right)^{\frac{d}{2}} + \left( \omega_{n}^2 \; + \; \mathcal{A}_2^2 \right)^{\frac{d}{2}} + \left( \omega_{n}^2 \; + \; \mathcal{A}_3^2 \right)^{\frac{d}{2}} + 3 \, \left( \omega_{n}^2 \; + \; \mathcal{A}_4^2 \right)^{\frac{d}{2}} \Bigg] \,.  
\end{equation}
The Matsubara sum can be written in terms of the following function~\cite{Chakrabortty:2024wto}
\begin{eqnarray}\label{eq:SQEDFTMS}
    \mathbb{S} [m^2] &=& \sum_{n}  \frac{\mu^{2 \epsilon}}{\beta ( 4 \pi)^{d/2}}  \; \bigg[ m^2  + \left( \frac{2\;n\; \pi}{\beta} \right)^2 \bigg]^{\frac{ d}{2}} \Gamma \left(\frac{ -d}{2} \right) \\
    &=& \frac{m^4}{32\pi^2} \Big(\log \Big[\frac{4\pi \mu^2}{m^2} \Big]+\frac{3}{2}-\gamma_E\Big) 
    + \frac{m^3}{6 \pi \beta} \; +\; \frac{\pi^2}{45\; \beta^4} - \frac{ m^2}{12\; \beta^2} \n \\ & & +\frac{m^4}{ (4\pi)^2}
\Big( \log \Big[\frac{m\beta e^{\gamma_E}}{4\pi} \Big] -\frac{3}{4}\Big )
 -\frac{2\zeta(3) \; m^6 \; \beta^2}{3 \; (4\pi)^4} +\frac{ \zeta (5) m^8 \beta ^4 }{ (4 \pi) ^6}+\cdots.   \nn
\end{eqnarray}
This leads to the effective potential in the Fermi gauge as
\begin{eqnarray}
    V_{\text{CW}}^{\beta}|^{{\rm FG}} ( \hat{\phi} ) &=& - \frac{1}{2} \; \bigg[\mathbb{S}[\mathcal{A}_1^2] + \mathbb{S}[\mathcal{A}_2^2] + \mathbb{S}[\mathcal{A}_3^2] + 3 \, \mathbb{S}[\mathcal{A}_4^2] \bigg] \n \\
   &=& -\frac{\pi^2}{15 \beta^4} +\frac{(9e^2+2 \lambda) \hat \phi^2}{72 \beta^2} -\frac{e^3 \hat \phi^3}{4 \pi \beta} \n \\
& & 	+\frac{\sqrt{3} \hat \phi^3}{864 \pi \beta}\,\Bigg[\sqrt {\lambda^2 -24 e^2 \xi \lambda} 
	 \left(\sqrt{\lambda-\sqrt{\lambda^2 -24 e^2 \xi \lambda}} 
    -\sqrt{\lambda+\sqrt{\lambda^2 -24 e^2 \xi \lambda}}\right) \Bigg] \n \\
   & &   - \frac{\sqrt{2} \lambda \hat \phi^3}{48 \pi \beta} \sqrt{\lambda} -\frac{\sqrt{3}  \lambda \hat \phi^3}{864 \pi \beta} \Bigg[   \left( \sqrt{\lambda-\sqrt{\lambda^2 -24 e^2 \xi \lambda}} 
      + \sqrt{\lambda+\sqrt{\lambda^2 -24 e^2 \xi \lambda}} \right) \Bigg]  \\
  & &    + \frac{\hat \phi^4}{64 \pi^2} \left(3 e^4 + \frac{5}{18} \lambda^2 - \frac{1}{3} e^2 \lambda \xi \right)
    {\rm log} \left( \frac{\hat{\phi}^2}{\mu^2}\right) 
     - \frac{\hat \phi^4}{32 \pi^2} \left(3 e^4 + \frac{5}{18} \lambda^2 - \frac{1}{3} e^2 \lambda \xi \right)
     {\rm log}\left( \frac{\hat \phi \beta e^{\gamma_E}}{4 \pi}\right) \nonumber \\ 
     && + \frac{\beta^2 \hat \phi^6}{256 \pi^4} \left( e^6 + \frac{7 \lambda^3}{162} - \frac{1}{36} e^2 \lambda^2 \xi \right) \zeta(3) +\cdots \,,\n
\end{eqnarray}
with a highly non-trivial gauge-dependence. In the Landau gauge $\xi=0$, the effective potential at finite temperature is therefore
\begin{eqnarray}\label{eq:Veff-FT-LG}
    V_{\text{CW}}^{\beta}|^{\rm LG} &=& -\frac{\pi^2}{15 \beta^4} +\frac{(9e^2+2 \lambda) \hat \phi^2}{72 \beta^2} -\hat \phi^3 \left( \frac{(9 + \sqrt{3}) \lambda^{3/2} + 54 \sqrt{2} e^3}{216 \sqrt{2} \pi \beta} \right) \nonumber \\
  & &    + \frac{\hat \phi^4}{64 \pi^2} \left(3 e^4 + \frac{5}{18} \lambda^2  \right)
     {\rm log} \left( \frac{\hat{\phi}^2}{\mu^2}\right)  \nonumber \\
    && - \frac{\hat \phi^4}{32 \pi^2} \left(3 e^4 + \frac{5}{18} \lambda^2  \right)
     {\rm log}\left( \frac{\hat \phi \beta e^{\gamma_E}}{4 \pi}\right)  + \frac{\beta^2 \hat \phi^6}{256 \pi^4} \left( e^6 + \frac{7 \lambda^3}{162} \right) \zeta(3) +\cdots .
\end{eqnarray}

\subsection{Effective Potential and Heat Kernel: Finite Temperature}
\label{subsec:FT-HK}
In case of finite temperature, the trace of the Heat Kernel is defined in a $(d+1)$-dimensional Euclidean manifold $(R^3 \times S^1)$ as \cite {Megias:2003ui, Moral-Gamez:2011wcb, Chakrabortty:2024wto}
\begin{eqnarray}\label{eq:TrHK}
    \text{\rm tr} \; K(t; x, x; \Delta) & =& \text{\rm tr} \int \frac{d^{d + 1} p}{(2 \pi)^{d + 1}} e^{- M^2 t} \langle x | \text{\rm exp} \bigg[- \int_0^t \left(- D^2 +U \right) dt \bigg] |p \rangle \langle p | x \rangle \nonumber \\
     &=& \frac{1}{\beta} \sum_{p_0} \int \frac{d^d p}{(2 \pi)^d} e^{-M^2 t}\text{\rm exp} \bigg[- \left[- \left(D_i + i p_i \right)^2 - Q^2 +U \right] t \bigg],
\end{eqnarray}
where the same is defined for zero temperature in Eq.~\ref{eq:ZHK}.\footnote{In this computation, all the masses are $\mathcal{O}(\hat{\phi}^2)$, and thus assumed to be nearly degenerate. In that consideration, we ignore the mixing effects due to non-degenerate masses, which, in any case, will not alter our conclusion. Thus, that effect is safely discarded without loss of generality.} Here, $Q=D_0+ip_0 \equiv \partial_0 + i \omega_n$, with $\omega_n={2\beta^{-1}n\pi}$ is the Matsubara frequency. The strong elliptic operator $\Delta =D^2+M^2+U$ reads as $\mathcal{D}$ in Eq.~\ref{eq:GreenFunction}. The $M^2$ and $U$ matrices are identified as
\begin{equation}\label{eq:FT-MU}
    M^2 = \begin{pmatrix}
        \frac{\lambda}{6}\hat{\phi}^2\, \delta_{ab} + \frac{\lambda}{3}\hat{\phi}_a \hat{\phi}_b & \quad 0 \\
        0  & e^2 \hat{\phi}^2\, \delta_{\mu \nu}
    \end{pmatrix},
    \quad {\rm and} \quad
    U = \begin{pmatrix}
        0 & e\, \epsilon_{ab'}\, \partial_\mu \hat{\phi}_{b'}  \\
        -e\, \epsilon_{a'b}\, \partial_\nu \hat{\phi}_{a'} & \, \left(1- {\xi}^{-1}\right)\partial_\mu \partial_\nu  
    \end{pmatrix}\,.
\end{equation}
Similarly to the previous section, we work in the temporal gauge $A_0=0$.  Following the prescription of Ref.~\cite{Chakrabortty:2024wto}, the trace of the HK can be expressed as 
\begin{eqnarray}\label{Kxx3}
	\text{tr} K (t;x,x;\Delta) & = &   \frac{1}{\beta}\; {\rm tr}   \sum_{p_0} \frac{1}{(4\pi t)^{\frac{d}{2}}} e^{-M^2t} e^{Q^2t} 
	[ \tilde{b}_0 - \tilde{b}_1 t +  \tilde{b}_2 t^2/2! + \cdots] \nonumber \\ 
    	&=&  \frac{1}{(4\pi t)^{\frac{d+1}{2}}}\; {\rm tr}\; \Bigg[e^{-M^2t} \Big[ \tilde{b}_0 \Theta_0 - \tilde{b}_1 t \Theta_0 +\cdots \Big] \Bigg]\,,
\end{eqnarray}    
where \begin{equation}
    \Theta_k(\Omega; t/\beta^2) = (4\pi t)^{1/2} \frac{1}{\beta}
    \sum_{p_0=\frac{2\pi n }{\beta}} t^{k/2} Q^k e^{Q^2 t},
\end{equation}
is the thermal wave function~\cite{Chakrabortty:2024wto}, and $\tilde{b}_i$'s are the finite temperature HK coefficients, with $\tilde{b}_0=b_0=\mathbb{1}$~\cite{Chakrabortty:2024wto}.
The one-loop effective Lagrangian at finite temperature is given as 
\begin{eqnarray}\label{eq:eft_lag_finite_temp}
	\mathcal{L}_{\text{eff}} &=& \frac{1}{2}\;\; \text{tr}\; \int_0^{\infty} \frac{dt}{t} K  
    	 = \frac{1}{2} \;\; \text{tr}\; \Bigg[  \tilde{b}_0 \mathbb{I}[0;0](M^2) - \tilde{b}_1 \mathbb{I}[0;1](M^2) + \cdots \Bigg],
\end{eqnarray}
where
\bea
\mathbb{I} [k;l] (M^2) &=&\sum_{n}  \left[\frac{2\pi  \iu }{\beta}\right]^k \frac{\mu^{2\epsilon}}{\beta (4\pi)^{d/2}} 
	[n]^k \; \left[ M^2+ \left(\frac{2 n \pi}{\beta}\right)^2 \right]^{-\frac{2l+k-d}{2}}  
	\Gamma \left(\frac{2l+k-d}{2}\right)\,.
\eea
The effective potential arises only from the first term; thus, at finite temperature, it is given as
\begin{eqnarray}
     V_{\text{CW}}^{\beta} ( \hat{\phi} ) &=& - \frac{1}{2}\;\; \text{tr}\; \Big[  \tilde{b}_0 \mathbb{I}[0;0] (M^2) \Big] \n \\
      &=& -\frac{6\pi^4}{90 \beta^4}  + \frac{  M_1^2+M_2^2+3M_3^2 }{24 \beta^2} -\frac{M_1^3+M_2^3+3M_3^3}{12 \pi \beta} \nonumber \\
    && + \frac{1}{64 \pi^2} \Bigg[ M_1^4 \log \left( \frac{ M_1^2}{\mu^2} \right) +   M_2^4 \log  \left(\frac{M_2^2}{\mu^2} \right)  + 3 M_3^4 \log \left( \frac{ M_3^2}{\mu^2} \right) \Bigg] \nonumber \\
    && -\frac{1}{32 \pi^2} \bigg[ M_1^4 \log \left( \frac{M_1 \beta e^{\gamma_E}}{4\pi} \right) +  M_2^4 \log \left(  \frac{M_2 \beta e^{\gamma_E}}{4\pi}  \right) + 3 M_3^4 \log \left( \frac{M_3 \beta e^{\gamma_E}}{4\pi} \right) \bigg] \nonumber \\
    && +\frac{\zeta(3) \beta^2}{3 (4 \pi)^4} \left( M_1^6 + M_2^6 + 3 M_3^6\right) +\cdots \nonumber \\
    &=& -\frac{\pi^4}{15 \beta^4} + \frac{(9 e^2 + 2 \lambda)}{72 \beta^2} \hat \phi^2 - \hat \phi^3 \left( \frac{(9 + \sqrt{3}) \lambda^{3 / 2} \; + \; 54\sqrt{2} e^3}{216 \sqrt{2} \pi \beta} \right) \nonumber \\
    && + \frac{\hat \phi^4}{64 \pi^2} \left(\frac{5 \lambda^2}{18} + 3 e^4 \right)  \log \left(\frac{\hat \phi^2}{\mu^2} \right)   \nonumber \\
    && - \frac{\hat \phi^4}{32 \pi^2} \left(\frac{5 \lambda^2}{18} + 3 e^4 \right) \log \left( \frac{\hat \phi \beta e^{\gamma_E}}{4 \pi}\right)  + \frac{\beta^2 \hat \phi^6}{256 \pi^4} \left( e^6 + \frac{7 \lambda^3}{162}\right) \zeta(3) +\cdots ,
\end{eqnarray}
where $M_i^2$ are the eigenvalues of the mass matrix $M^2$, see Eq.~\ref{eq:FT-MU}, $ M_1^2 = \frac{\lambda \hat{\phi}^2}{6} \; ; \; M_2^2 = \frac{\lambda \hat{\phi}^2}{2} \; ; \; M_3^2 = e^2 \hat{\phi}^2$, and $M_i^3 \equiv (M_i^2)^{3/2}$. The potential, computed using the HK prescription, is gauge independent and exactly matches the same computed in the functional method in the Landau gauge; see Eq.~\ref{eq:Veff-FT-LG}.
   
\section{Conclusions }
\label{sec:conclusion}
%
The consideration of the effective potential is ubiquitous in high-energy phenomenology. It provides a well-motivated avenue for clarifying the vacuum structure of a quantum field theory, and it is therefore a central object in the study of early universe phase transitions and inflationary cosmology. A drawback arises when the effective potential exhibits non-vanishing gauge-dependence. In practical calculations, therefore, the freedom provided by the Nielsen identities is then exploited to remove gauge-dependencies through redundant field redefinitions, which can then provide the starting point of a more comprehensive study of gauge-robustness (see, e.g.~\cite{Patel:2011th,Espinosa:2015qea}). If this is the strategy of choice, a more practical approach is to avoid gauge parameter dependencies from the beginning. Extending the results of~\cite{Balui:2025kat} to the broken phase, in this work we have traced their spurious origin to an incomplete factorisation of elliptic operator traces that can be mended in $d=4$ by including a multiplicative anomaly. It is demonstrated, first for the Abelian Higgs model and then for the Standard Model, that the inclusion of these terms in the Fermi gauge exactly cancels the `naive' gauge-dependence of the potential that would otherwise need removing through a field redefinition. In a sense, the inclusion of the anomaly is interchangeable with a field redefinition via the Nielsen identity in practical terms.
Infrared divergences in theories with spontaneously broken symmetries that generically require resummation~\cite{Espinosa:2015qea, Espinosa:2016uaw} are further mended when including the anomaly in the Fermi gauge, which then yields a result consistent with the Landau gauge computation. Gauge dependencies and IR-singular behaviour are addressed simultaneously via the multiplicative anomaly. Furthermore, we have shown that these technical obstacles can be circumvented by utilising the Heat Kernel method. The latter, therefore, provides the most economical approach to obtaining robust results (of course, within the premise of our assumptions).  

Finally, we have demonstrated that this improved behaviour extends to the finite temperature domain of the effective potential. Specifically employing the Heat Kernel method, the finite temperature effective potential in gauge theories is found to be gauge-invariant at the considered order. In contrast, the standard approach yields a finite temperature effective potential that depends on the gauge choice, just as the zero temperature potential does. On the one hand, the Nielsen identities can be used to shift the gauge-dependence to higher orders in the coupling expansion~\cite{Patel:2011th, Andreassen:2014eha, Urbano:2019ohp, Garny:2012cg}. On the other hand, the one-loop effective potential resums coupling orders consistently and is equally valid for strongly coupled theories. Therefore, the gauge independence in the leading-order perturbative expansion weakens the validity of applications. In this paper, in contrast, we construct pathways for explicit gauge independence of the effective potential at both zero and finite temperatures, valid at all orders in the coupling constant and for all field values of the potential. This advances the use of effective potentials in phase transitions, inflation, stochastic wave computations, and baryogenesis, serving both as a theoretical demonstration of targeted computations in effective field theories and as an extension of their range of applicability.

\subsection*{Acknowledgements}
We thank Jos\'e Espinosa, Mathias Garny, Thomas Konstandin, and Philipp Schicho for insightful discussions and comments.
DB, TB, JC, DD acknowledge support from the Science and Engineering Research Board (SERB), Government of India, under the Project SERB/PHY/2023799.
SM thanks the IIT Kanpur grant for a Distinguished Visiting Professorial position, which supports the implementation of this work.
CE is supported by the STFC under grant ST/X000605/1, and by the Leverhulme Trust under Research Fellowship RF-2024-300$\backslash$9. C.E. is further supported by the Institute for Particle Physics Phenomenology Associateship Scheme.

\appendix	
\section{Standard Model: Effective Potential}
\label{sec:SMEP}
Extending the Abelian Higgs model to the more realistic case of the Standard Model (SM), we turn to the SM Lagrangian~\cite{DiLuzio:2014bua,Espinosa:2016nld, Espinosa:2016uaw}
\begin{equation} \label{eq:SMLag1}
\mathcal{L}_{\rm{SM}} = \mathcal{L}_{\rm{YM}} + \mathcal{L}_{\rm{H}} + \mathcal{L}_{\rm{Y}} + \mathcal{L}^{\rm{Fermi}}_{\rm{g.f.}} \,  
\end{equation} 
Here, we focus on the Lagrangian of the electroweak sector $(SU(2)_L \times U(1)_Y)$, which is relevant in effective potential computation:
\begin{eqnarray}\label{eq:SMLag2}
\mathcal{L}_{\rm{YM}} &= & 
-\frac{1}{4} \left( \partial_\mu W^a_\nu - \partial_\nu W^a_\mu  + g \epsilon^{abc} W^b_\mu W^c_\nu \right)^2
-\frac{1}{4} \left( \partial_\mu B_\nu - \partial_\nu B_\mu  \right)^2  , \n \\
\mathcal{L}_{\rm{H}} &=& \left( D_\mu H \right)^\dagger \left( D^\mu H \right) - V(H)  , \n \\
\mathcal{L}_{\rm{Y}} &=& \overline{Q}_L i \gamma_\mu D^\mu Q_L + \overline{t}_R i \gamma_\mu D^\mu t_R + 
\left(- y_t \overline{Q}_L (i\sigma^2) H^* t_R + \text{h.c.}\right)   \,, \n \\
\mathcal{L}^{\rm{Fermi}}_{\rm{g.f.}} & = & -\frac{1}{2 \xi_W} \left( \partial^\mu W^a_\mu \right)^2 
-\frac{1}{2 \xi_B} \left( \partial^\mu B_\mu \right)^2  \,, 
\end{eqnarray}
where the covariant derivative\footnote{In case of $SU(2)_L$ singlet, i.e., right-handed fields, covariant derivative reads as $D_\mu \equiv \partial_\mu  + i g' \frac{Y}{2} B_\mu$.} is $D_\mu \equiv \partial_\mu - i g \frac{\sigma^a}{2} W^a_\mu + i g' \frac{Y}{2} B_\mu$, and potential $V (H) = -m^2 H^\dag H + \lambda (H^\dag H)^2 $. Here, $Q_L = (t_L, b_L)^T$, and $t_R$ are the $SU(2)_L$ quark (third-generation) doublet and singlet, respectively. Among fermions, we retain only the top quark, whose large Yukawa coupling gives the dominant contribution to the effective potential. Our results can be straightforwardly generalised to other fermion species. For the sake of this work, we suppress other internal symmetry indices. In the broken phase, the SM Higgs doublet is given by \cite{DiLuzio:2014bua,Espinosa:2016nld, Espinosa:2016uaw}
\begin{equation}
H(x) \equiv
\frac{1}{\sqrt{2}}
\left( 
\begin{array}{c}
\chi_1(x) + i \chi_2(x) \\
\phi + h(x) + i \chi_3(x)
\end{array} 
\right)\,,  
\end{equation}
where $h$ and $\chi_a$ ($a=1,2,3$) denote the Higgs and Goldstone bosons, respectively, and $\phi$ is the constant background.

The one-loop effective potential, after integrating out bosonic and fermionic fluctuations, can be written as
\begin{eqnarray} \label{eq:SMVeff}
V^{(1)}_{\rm{eff}} (\phi)& = & i\sum_{n \, 
\in \, \text{SM}} c_s \int \frac{d^4 k}{(2\pi)^4} \log \det i \tilde{\mathcal{G}}^{-1}_n \{ \phi; k \} \,.
\end{eqnarray}
where $c_s=-1/2$, and $1$ for real-bosonic and fermion fields, respectively. In the broken phase, the physical masses of the SM particles are given as a function of the background $\phi$ as \cite{DiLuzio:2014bua,Espinosa:2016nld, Espinosa:2016uaw}
\begin{eqnarray}
M_h^2 &=& -m^2 + 3 \lambda \phi^2 \,, 
~M_{\chi}^2 = -m^2 + \lambda \phi^2  \,, \nonumber\\
M_W^2 &=& \tfrac{1}{4} g^2 \phi^2 \,, ~
M_B^2 = \tfrac{1}{4} g'^{2} \phi^2 \,,~M_t^2 = \frac{y_t^2}{2} \phi^2  \,,
\end{eqnarray}
where $M_W^2=M_{W^i}^2 ,\;\; M_\chi^2=M_{\chi_i}^2$ $\forall \; i\in\{1,2,3\}$.
To compute the various $\tilde{\mathcal{G}}^{-1}_n$ required for Eq.~\ref{eq:SMVeff}, we start with the bosonic degrees of freedom
\begin{equation}
\label{quadgoldgauge}
\frac{1}{2} X^{T} \left( i \mathcal{G}_X^{-1} \right) X
= \frac{1}{2}
(V^T_\mu,\chi^T , h )
^T
\begin{pmatrix}
    i \left( \mathcal{G}_V^{-1} \right)^{\mu}{}_{\nu} & \mathcal{M}_{\text{mix}}^{T}\,\partial^{\mu} & 0 \\
    - \mathcal{M}_{\text{mix}}\,\partial_{\nu} & i \mathcal{G}_\chi^{-1} & 0 \\
    0 & 0 & i \mathcal{G}_h^{-1}
\end{pmatrix}
\begin{pmatrix}
    V^{\nu} \\
    \chi \\
    h
\end{pmatrix}
\,,
\end{equation}
where we have introduced
\begin{equation}
\label{extX}
X^T = \left(V_\mu^T , \chi^T , h\right) \,,
\end{equation}
with
\begin{equation}
\label{defVchiFermi}
V^T_\mu = \left( W^1_\mu, W^2_\mu, W^3_\mu, B_\mu \right)  ~
\text{and}~\chi^T = \left( \chi_1, \chi_2, \chi_3 \right)  \,, 
\end{equation}
as well as
\begin{equation}
\label{defmmix}
\mathcal{M}_{\text{mix}} = 
\begin{pmatrix}
    0       & -M_W   & 0     & 0 \\
    -M_W    & 0      & 0     & 0 \\
    0       & 0      & M_W   & M_B
\end{pmatrix}
\,.
\end{equation}
The gauge boson and Goldstone Green's functions in momentum space are
\begin{equation}
\label{invpropgaugeFermi}
i \tensor{( \tilde{\mathcal{G}}^{-1}_V )}{^\mu_\nu}  =
i \tilde{\mathcal{G}}^{-1}_{T} \,\left(\tensor{g}{^\mu_\nu} - \frac{k^\mu k_\nu}{k^2}\right)
+ i \tilde{\mathcal{G}}^{-1}_{L} \, \left(\frac{k^\mu k_\nu}{k^2} \right)\,, 
\end{equation}
\begin{equation}
\label{invpropTFermi}
i \tilde{\mathcal{G}}^{-1}_{T} = 
\begin{pmatrix}
    - k^2 + M_W^2 & 0 & 0 & 0 \\
    0 & - k^2 + M_W^2 & 0 & 0 \\
    0 & 0 & - k^2 + M_W^2 & M_W M_B \\
    0 & 0 & M_W M_B & - k^2 + M_B^2
\end{pmatrix}
\,,
\end{equation}
\begin{equation}
\label{invpropLFermi}
i \tilde{\mathcal{G}}^{-1}_{L} = 
\begin{pmatrix}
    - \xi_W^{-1} k^2 + M_W^2  & 0 & 0 & 0 \\
    0 & - \xi_W^{-1} k^2 + M_W^2 & 0 & 0 \\
    0 & 0 & - \xi_W^{-1} k^2 + M_W^2  & M_W M_B \\
    0 & 0 & M_W M_B & - \xi_B^{-1} k^2 + M_B^2
\end{pmatrix}
\,,
\end{equation}
\begin{equation}
\label{invpropchiFermi}
i \tilde{\mathcal{G}}^{-1}_{\chi} = 
\begin{pmatrix}
    k^2 - M_\chi^2 & 0 & 0 \\
    0 & k^2 - M_\chi^2 & 0 \\
    0 & 0 & k^2 - M_\chi^2
\end{pmatrix}
\,.
\end{equation}
Finally, the Higgs and top quark Green's functions are given as
\begin{eqnarray} \label{eq:SMHtopProp}
    i \tilde{\mathcal{G}}^{-1}_h &=& k^2 - M_h^2\,, \;\;\;\;\;
    i \tilde{\mathcal{G}}^{-1}_t = \slashed{k} - M_t\, ,
\end{eqnarray}
respectively.
In dimensional regularisation $(d=4-2\epsilon)$, the one-loop effective potential (OLEP) is given as \cite{DiLuzio:2014bua,Espinosa:2016nld, Espinosa:2016uaw}
\begin{eqnarray}
\label{EP1loopSMdrexplFermi2}
V^{(1)}_{\rm{eff}}|^{\rm{F}}_{\rm SM}  &=& -\frac{i}{2} \mu^{2\epsilon} \int \frac{d^d k}{(2\pi)^d} 
\left[
- 12 \log \left( - k^2 + M_t^2 \right) + (d-1) \left( 2 \log \left( -k^2 + M_W^2 \right) 
\right. \right. \nonumber \\ 
&& \left. + \log \left( -k^2 + M_Z^2 \right) \right) + \log \left( k^2 - M_h^2 \right) 
+ 2 \log \left( k^2 - M_{A_{+}}^2 \right) + 2 \log \left( k^2 - M_{A_{-}}^2 \right)  
\nonumber \\ 
& &\left. + \log \left( k^2 - M_{B_{+}}^2 \right) + \log \left( k^2 - M_{B_{-}}^2 \right) + \cdots \right]  ,
\end{eqnarray} 
where 
\begin{eqnarray} 
\label{defmassZFermi2}
M_{Z}^2 &=& M_W^2 + M_B^2  , 
{M}_{A_{\pm}}^2 = \frac{1}{2}  \left(  M_\chi^2 \pm \sqrt{ M_\chi^4 - 4 \xi_W M_\chi^2 M_W^2} \right)  \,, \\ 
{M}_{B_{\pm}}^2 &=& \frac{1}{2} \left(  M_\chi^2 \pm \sqrt{ M_\chi^4
- 4 M_\chi^2 (\xi_W M_W^2 + \xi_B M_B^2) } \right)  \,. 
\end{eqnarray} 
Employing the $\overline{\text{MS}}$ renormalisation scheme, the OLEP can be written as \cite{DiLuzio:2014bua,Espinosa:2016nld, Espinosa:2016uaw}
\begin{eqnarray}\label{eq:SMVeff1L} 
 V^{(1)}_{\rm{eff}}|^{\rm{F}}_{\rm SM} &=& \frac{1}{64 \pi^2} \left[ 
-12 M_t^4 \left( \log\frac{M_t^2}{\mu^2} - \frac{3}{2} \right)
+6 M_W^4 \left( \log\frac{M_W^2}{\mu^2} - \frac{5}{6} \right) \right.\nonumber  \\
&& \left.+3 M_Z^4 \left( \log\frac{M_Z^2}{\mu^2} - \frac{5}{6} \right) 
+M_h^4 \left( \log\frac{M_h^2}{\mu^2} - \frac{3}{2} \right)
+2 M_{A_+}^4 \left( \log\frac{M_{A_+}^2}{\mu^2} - \frac{3}{2} \right)  \right. \\
&& \left. +2 M_{A_-}^4 \left( \log\frac{M_{A_-}^2}{\mu^2} - \frac{3}{2} \right) 
+ M_{B_+}^4 \left( \log\frac{M_{B_+}^2}{\mu^2} - \frac{3}{2} \right)
+ M_{B_-}^4 \left( \log\frac{M_{B_-}^2}{\mu^2} - \frac{3}{2} \right)
\right]  \,. \nonumber
\end{eqnarray}

\subsection{Landau Gauge}
\label{subsec:LGSM}
As in our discussion of the Abelian Higgs model, we set the stage by considering the Landau gauge, $\xi=0$. In this case, the OLEP Eq.~\ref{eq:SMVeff1L}, reads~\cite{Martin:2014bca,Elias-Miro:2014pca,DiLuzio:2014bua,Espinosa:2016nld, Espinosa:2016uaw}
\begin{eqnarray}\label{eq:SMVeff-LG}
    V^{(1)}_{\rm eff}\big|^{\rm L}_{\rm SM} &=& \frac{1}{64 \pi^2} \bigg[ 
                -12 M_t^4 \left( \log\frac{M_t^2}{\mu^2} - \frac{3}{2} \right)
                + 6 M_W^4 \left( \log\frac{M_W^2}{\mu^2} - \frac{5}{6} \right)
                + 3 M_Z^4 \left( \log\frac{M_Z^2}{\mu^2} - \frac{5}{6} \right) \nonumber \\
         &   &\quad + M_h^4 \left( \log\frac{M_h^2}{\mu^2} - \frac{3}{2} \right)
                + 3  M_\chi^4 \left( \log  \frac{M_\chi^2}{\mu^2} - \frac{3}{2}\right) 
            \bigg]\,.
\end{eqnarray}

\subsection{Fermi Gauge}
\label{subsec:FGSM}
Moving on to the Fermi gauge, ($\xi \neq 0$), we again assume $4 \xi_W M_W^2  \gg  M_\chi^2$ and $4 \left( \xi_W M_W^2 + \xi_B M_B^2 \right)  \gg  M_\chi^2$ that is consistent with the vanishing Goldstone mass limit in the spirit of Sec.~\ref{subsec:FGSQED}. In this limit, the Goldstone masses read as  
 \begin{eqnarray}
       M_{A_\pm}^2 &\simeq& \frac{M_\chi^2}{2} \pm i\sqrt{\xi_W M_W^2 M_\chi^2} \,, \\
        M_{B_\pm}^2 &\simeq& \frac{M_\chi^2}{2} \pm i\sqrt{ \left( { \xi_W M_W^2 + \xi_B M_B^2} \right) M_\chi^2 }\,.
 \end{eqnarray}
The Goldstone mass dependent term in the effective potential in the Fermi gauge, therefore, is
\begin{eqnarray}\label{eq:SMVeffGFG}
 V^{(1)}_{\rm eff} [M_\chi^2]
&=& \frac{1}{64 \pi^2} \Bigg[ 2 M_{A_+}^4 \left( \log \frac{ M_{A_+}^2}{\mu^2} - \frac{3}{2} \right)  +  2 M_{A_-}^4 \left( \log \frac{M_{A_-}^2}{\mu^2} - \frac{3}{2} \right) \nonumber \\
&& + M_{B_+}^4 \left( \log \frac{M_{B_+}^2}{\mu^2} - \frac{3}{2} \right)  +  M_{B_-}^4 \left( \log \frac{M_{B_-}^2}{\mu^2} - \frac{3}{2} \right) \Bigg] \nonumber \\
&\simeq& \frac{1}{64 \pi^2}\Bigg[ 4 \left( \frac{M_\chi^4}{4} - \xi_W M_\chi^2 M_W^2\right) \left( \log \frac{|M_{A_+}^2|}{\mu^2} -\frac{3}{2} \right) \nonumber \\
&& + 2 \left( \frac{M_\chi^4}{4} - M_\chi^2 (\xi_W M_W^2 + \xi_B M_B^2) \right) \left( \log  \frac{|M_{B_+}^2|}{\mu^2}  -\frac{3}{2} \right)\Bigg]\,.
 \end{eqnarray}
The full OLEP in the Fermi gauge is expressed as \cite{DiLuzio:2014bua,Espinosa:2016nld, Espinosa:2016uaw}
        \begin{eqnarray}\label{eq:SMVeffFG}
            \tilde{V}^{(1)}_{\rm eff}\big|^{\rm F}_{\rm SM} &=& \frac{1}{64\pi^2} \Bigg[ 
                -12 M_t^4 \left( \log\frac{M_t^2}{\mu^2} - \frac{3}{2} \right)
                + 6 M_W^4 \left( \log\frac{M_W^2}{\mu^2} - \frac{5}{6} \right)
                + 3 M_Z^4 \left( \log\frac{M_Z^2}{\mu^2} - \frac{5}{6} \right) \nonumber \\
            & & + M_h^4 \left( \log\frac{M_h^2}{\mu^2} - \frac{3}{2} \right)
                + 4 \left( \frac{M_\chi^4}{4} - M_\chi^2 \xi_W M_W^2 \right) \left( \log  \frac{|M_{A_+}^2|}{\mu^2}  -\frac{3}{2} \right)\nonumber \\
            & & + 2 \left( \frac{M_\chi^4}{4} - M_\chi^2 (\xi_W M_W^2 + \xi_B M_B^2) \right) \left( \log  \frac{|M_{B_+}^2|}{\mu^2} -\frac{3}{2} \right)
            \Bigg]\,.
\end{eqnarray}

\subsection{Fermi Gauge and the Multiplicative Anomaly}
\label{subsec:FGMASM}
The $\xi$-dependent multiplicative anomaly in the case of the SM is given as, following Eq.~\ref{eq:SQEDMA}
\begin{eqnarray}\label{eq:SMMA} 
	\mathbb{a1}[M_{A_+}^2, M_{A_-}^2] (\xi)=	\frac{\mathbb{A}[M_{A_+}^2, M_{A_-}^2] (\xi)}{\mathcal{V}_4} &=&  \frac{1}{64 \pi^2}  \left(-4 \xi_W M_W^2 M_\chi^2 \right) \\
    	\mathbb{a2}[M_{B_+}^2, M_{B_-}^2] (\xi)=	\frac{\mathbb{A}[M_{B_+}^2, M_{B_-}^2] (\xi)}{\mathcal{V}_4} 
&=& \frac{1}{64 \pi^2}  \left(-4 (\xi_W M_W^2 + \xi_B M_B^2) M_\chi^2 \right)\,.
\end{eqnarray}
The anomaly-corrected OLEP in the Fermi gauge is
\begin{eqnarray}
\label{1loopEPbareFermi}\label{eq:SMVeffFGMA}
V^{(1)}_{\rm{eff}}|^{\rm F +MA}_{\rm SM} &=&  \tilde{V}^{(1)}_{\rm eff}\big|^{\rm F}_{\rm SM}  - \frac{1}{2} \; \times 2\;\mathbb{a1}[M_{A_+}^2, M_{A_-}^2](\xi) \left( \log \frac{|M_{A_+}^2|}{\mu^2}  - \frac{3}{2} \right)  \nonumber\\ & & \nonumber\hspace{6cm}- \frac{1}{2} \; \mathbb{a2}[M_{B_+}^2, M_{B_-}^2](\xi) \left( \log \frac{|M_{B_+}^2|}{\mu^2}  - \frac{3}{2} \right)\nonumber \\
&=&\frac{1}{64\pi^2} \Bigg[ 
                -12 M_t^4 \left( \log\frac{M_t^2}{\mu^2} - \frac{3}{2} \right)
                + 6 M_W^4 \left( \log\frac{M_W^2}{\mu^2} - \frac{5}{6} \right)
                + 3 M_Z^4 \left( \log\frac{M_Z^2}{\mu^2} - \frac{5}{6} \right) \nonumber \\
            && \hspace{1cm} + M_h^4 \left( \log\frac{M_h^2}{\mu^2} - \frac{3}{2} \right)
                + 4 \left( \frac{M_\chi^4}{4} - M_\chi^2 \xi_W M_W^2 \right) \left( \log  \frac{|M_{A_+}^2|}{\mu^2}  -\frac{3}{2} \right)\nonumber \\
            && \hspace{1cm} + 2 \left( \frac{M_\chi^4}{4} - M_\chi^2 (\xi_W M_W^2 + \xi_B M_B^2) \right) \left( \log  \frac{|M_{B_+}^2|}{\mu^2} -\frac{3}{2} \right)
            \Bigg] \nonumber \\
            && - \frac{1}{2} \; \times 2\;\mathbb{a1}[M_{A_+}^2, M_{A_-}^2](\xi) \left( \log \frac{|M_{A_+}^2|}{\mu^2} - \frac{3}{2} \right)  \nonumber\\ & &\nonumber\hspace{6cm} - \frac{1}{2} \mathbb{a2}[M_{B_+}^2, M_{B_-}^2](\xi) \left( \log \frac{|M_{B_+}^2|}{\mu^2}  - \frac{3}{2} \right) \nonumber \\
&\simeq& \frac{1}{64 \pi^2} \Bigg[
-12 M_t^4 \left( \log\frac{M_t^2}{\mu^2} - \frac{3}{2} \right) 
+ 6 M_W^4 \left( \log\frac{M_W^2}{\mu^2} - \frac{5}{6} \right) +3 M_Z^4 \left( \log\frac{M_Z^2}{\mu^2} - \frac{5}{6} \right)  \nonumber \\
&&+M_h^4 \left( \log\frac{M_h^2}{\mu^2} - \frac{3}{2} \right) + \frac{3}{2} M_\chi^4 \left( \frac{1}{2}\log  \frac{M_{\chi}^2}{\mu^2}  - \frac{3}{2} \right)
\Bigg]\,.  
\end{eqnarray}
Quite identical to our findings detailed in Sec.~\ref{subsec:NIMASQED}, the inclusion of the multiplicative anomaly addresses IR sensitivity and $\xi$-dependence simultaneously. The former is apparent by comparing the Landau result with Eqs.~\ref{eq:SMVeffFG} and \ref{1loopEPbareFermi}. Derivatives of the anomaly-corrected effective potential in the Fermi gauge show the improved IR behaviour apparent in the Landau gauge. In the next section, we investigate the $\xi$-dependence in conjunction with the Nielsen identity in more detail.

\subsection{Nielsen Identity with Multiplicative Anomaly}
\label{subsec:NIMASM}
We turn to the Nielsen identity \cite{Nielsen:1975fs} as in Eq.~\ref{eq:SQEDNI} by analysing the gauge-dependent part of the OLEP
\begin{multline}
    \tilde{V}^{(1)}_{\rm eff}|^{\rm F}_{\rm SM} (|M_{A_+}^2|, |M_{B_+}^2|) \\\equiv  - \frac{i}{2}\int \frac{d^4 k}{(2 \pi)^4} \bigg[ 2 \log \left[ (k^2-M_{A_+}^2) (k^2 - M_{A_-}^2) \right] + \log \left[ (k^2 - M_{B_+}^2) (k^2 - M_{B_-}^2) \right] \bigg]\,,
    \label{v1g_1}
\end{multline}
which yields
\begin{eqnarray}\label{eq:}
\sum_{i=W,B}\xi_i \frac{\partial \tilde{V}^{(1)}_{\rm eff}|^{\rm F}_{\rm SM}}{\partial \xi_i} &=& \frac{i}{2} \int \frac{d^4 k}{(2 \pi)^4} \Bigg[ 
2 \left( \xi_W \frac{\partial M_{A_+}^2}{\partial \xi_W} \frac{1}{(k^2 -M_{A_+}^2)} 
+ \xi_W \frac{\partial M_{A_-}^2}{\partial \xi_W} \frac{1}{(k^2 -M_{A_-}^2)} \right) \nonumber \\
&& + \left( \xi_B \frac{\partial M_{B_+}^2}{\partial \xi_B} \frac{1}{(k^2 -M_{B_+}^2)} 
+ \xi_B \frac{\partial M_{B_-}^2}{\partial \xi_B} \frac{1}{(k^2 -M_{B_-}^2)} \right) \nonumber \\
&& + \left( \xi_W \frac{\partial M_{B_+}^2}{\partial \xi_W} \frac{1}{(k^2 -M_{B_+}^2)} 
+ \xi_W \frac{\partial M_{B_-}^2}{\partial \xi_W} \frac{1}{(k^2 -M_{B_-}^2)} \right) 
\Bigg]\,.
\end{eqnarray}
 In this scenario, similar to the MSQED (see Sec.~\ref{subsec:NIMASQED}), we also focus on the terms that are of the form $\sim M_\chi^2 \log {M_\chi^2}/{\mu^2}$, which is relevant for the effective potential.

Therefore, we have
\begin{eqnarray}\label{eq:SMNIVderi}
\xi_W \frac{\partial \tilde{V}^{(1)}_{\rm eff}|^{\rm F}_{\rm SM}}{\partial \xi_W} + \xi_B \frac{\partial \tilde{V}_{\rm eff}^{(1)}|^{\rm F}_{\rm SM}}{\partial \xi_B} 
&=& \frac{1}{64 \pi^2} 
\bigg[(- 4 M_\chi^2 \xi_W M_W^2) \nonumber \\ 
&+& 
(- 2 M_\chi^2) \left( \xi_W M_W^2 + \xi_B M_B^2 \right)\bigg] \left( \frac{1}{2} \log \frac{M_\chi^2}{\mu^2} \right)\,.
\end{eqnarray}
The $\xi$-dependent part of the multiplicative anomaly is given as (see Eq.~\ref{eq:SMMA})
\begin{eqnarray}
    \mathbb{a1}[M_{A_+}^2, M_{A_-}^2] (\xi) &= & \frac{1}{64 \pi^2}  \left(-4 M_\chi^2 \xi_W M_W^2 \right), \nonumber \\
    \mathbb{a2}[M_{B_+}^2, M_{B_-}^2] (\xi) &= & \frac{1}{64 \pi^2}  \left(-4 M_\chi^2 \left( \xi_W M_W^2 + \xi_B M_B^2\right) \right)\,.
\end{eqnarray}
Thus, in the case of the SM, the Nielsen identity 
\begin{eqnarray}
    \sum_{i=W,B} \xi_i \frac{\partial \tilde{V}^{(1)}_{\rm eff}|^{\rm{F}}_{\rm{SM}}}{\partial \xi_i} - \frac{1}{2} \bigg[ \sum_{j=1,2} \;\;\sum_{i=W,B} \xi_i \frac{\partial \mathbb{a}_j (\xi_i)}{\partial \xi_i} \bigg] \left( \frac{1}{2} \log \frac{M_\chi^2}{\mu^2}\right) + \sum_{i=W,B} C^{(1)}_{i} \frac{\partial V^{(0)}}{\partial \phi} &=& 0\,,
\end{eqnarray}
leads to
\begin{multline}
     \Bigg[ -\frac{4 M_\chi^2 \xi_W M_W^2}{64 \pi^2}
    - \frac{2 M_\chi^2 \left(\xi_W M_W^2 + \xi_B M_B^2 \right)}{64 \pi^2} 
    - \frac{1}{2} \xi_W \left( 2 \cdot \frac{-4 M_\chi^2 M_W^2 }{64 \pi^2} \right)  \\
     - \frac{1}{2} \left( \xi_W \frac{- 4 M_\chi^2  M_W^2}{64 \pi^2} + \xi_B \frac{- 4 M_\chi^2 M_B^2}{64 \pi^2} \right) \Bigg] \left( \frac{1}{2}\log \frac{M_\chi^2}{\mu^2}\right)
    + \sum_{i=W,B} C^{(1)}_{i} \frac{\partial V^{(0)}}{\partial \phi} = 0 \,,
\end{multline}
from which follows
\begin{eqnarray}
\sum_{i=W,B} C^{(1)}_{i} \frac{\partial V^{(0)}}{\partial \phi} = 0\,.
\label{Nielsen_id_eq1_1}
\end{eqnarray}
Again, as the derivative is typically non-vanishing, the above equation implies that $C^{(1)}_{W} + C^{(1)}_{B} =0$ everywhere in the potential. Closer inspection tells us that $C^{(1)}_{W} = C^{(1)}_{B} =0$ as the gauge groups are independent. As for the Abelian Higgs model, once the contribution from the multiplicative anomaly is taken into account, the effective potential becomes gauge-independent, and the Nielsen identity is trivially fulfilled as a consequence. The conclusions and observations provided at the end of Sec.~\ref{subsec:NIMASQED}, therefore, directly generalise to the phenomenologically more interesting case of the SM: both gauge and IR dependencies upon including the multiplicative anomaly are simultaneously improved. 

\subsection{Heat Kernel Method}
\label{sec:HKSM}
We consider the same Standard Model Lagrangian, see Eqs.~\ref{eq:SMLag1}-\ref{eq:SMLag2} in Sec.~\ref{sec:SMEP}, to compute the effective Lagrangian 
\begin{equation}
\label{Lclassical}
\mathcal{L}_{\rm{SM}} = \mathcal{L}_{\rm{YM}} + \mathcal{L}_{\rm{H}} + \mathcal{L}_{\rm{f}} + 
\mathcal{L}^{\rm{Fermi}}_{\rm{g.f.}} \, .  
\end{equation} 
%
Here, the non-degenerate mass matrix in the basis $X = (W^a_\mu,B_\mu,\chi^i,h)^T$ with $a,i=\{1,2,3\}$, defined in Sec.~\ref{sec:SMEP}, is given as
\begin{eqnarray}
    \! \! \!  \! \! \! \! M^2 &=&
\begin{pmatrix}
 -(\tfrac{1}{2} g \phi)^2 \, \delta_{\mu \nu} I_{3\times3} & 0 & 0 & 0 \\
 0 & -(\tfrac{1}{2} g' \phi)^2 \, \delta_{\mu \nu} & 0  & 0 \\
0 & 0 & -m^2 + \lambda \phi^2 \, I_{3\times3} & 0 \\
0 & 0 & 0 &-m^2 + 3 \lambda \phi^2
\end{pmatrix}\,.
\end{eqnarray}
Following the HK method, see the previous section, we define the matrix $\mathcal{A}$ in the same basis $X$ as
\begin{multline}\label{eq:SM-A}
\mathcal{A} 
=\\
\resizebox{0.97\textwidth}{!}{$
\begin{pmatrix}
   \left[(\partial^2 \!+\! \frac{2ip\cdot\partial}{\sqrt{t}} ) \delta_{\mu \nu} \! + \! \left(1\!-\! \frac{1}{\xi_W}\right) \! \left(\partial_\mu \partial_\nu - \frac{p_\mu p_\nu}{t} + \frac{2ip_\mu \partial_\nu}{\sqrt{t}}\right)\right] I_{3\times3} & 0 & \left[g \partial_\mu \phi + \frac{ig p_\mu \phi}{\sqrt{t}}\right] I_{3\times1} &  0 \\
0 & \! \! (\partial^2 \!+\! \frac{2ip\cdot\partial}{\sqrt{t}} ) \delta_{\mu \nu} \! + \! \left(1\!-\! \frac{1}{\xi_B}\right) \! \left(\partial_\mu \partial_\nu - \frac{p_\mu p_\nu}{t} + \frac{2ip_\mu \partial_\nu}{\sqrt{t}}\right) & g' \partial_\mu \phi + \frac{ig' p_\mu \phi}{\sqrt{t}} &  0 \\
\! \left[-g \partial_\nu \phi - \frac{ig p_\nu \phi}{\sqrt{t}}\right] I_{1\times3} & \! -g' \partial_\nu \phi - \frac{ig' p_\nu \phi}{\sqrt{t}} &  \left[\partial^2 \!+\! \frac{2ip\cdot\partial}{\sqrt{t}}\right] I_{3\times3} & 0\\
  0  &  0  &  0 &  \partial^2 \!+\! \frac{2ip\cdot\partial}{\sqrt{t}}
\end{pmatrix}
$} \,.
\end{multline}
After incorporating $\mathcal{A}$, Eq.~\ref{eq:SM-A}, in the effective Lagrangian, Eq.~\ref{eq:HKLeff}, we find the one-loop effective potential as
\begin{eqnarray}\label{eq:SMVeff-HK} 
 V^{(1)}_{\rm{eff}}|^{\rm{HK}}_{\rm SM} &=& \frac{1}{64 \pi^2} \text{tr} \Bigg[ 
- M_t^4 \left( \log  \frac{M_t^2}{\mu^2} - \frac{3}{2} \right)
+ M_W^4 \left( \log \frac{M_W^2}{\mu^2} - \frac{5}{6} \right)  \\
&+& {M}_Z^4 \left( \log \frac{{M}_Z^2}{\mu^2}   - \frac{5}{6} \right) 
+M_h^4 \left( \log  \frac{M_h^2}{\mu^2}  - \frac{3}{2} \right)
+M_{\chi}^4 \left( \log  \frac{M_{\chi}^2}{\mu^2} - \frac{3}{2} \right) \Bigg] \nonumber \\
&=&  \frac{1}{64 \pi^2} \Bigg[ 
- 12 \, M_t^4 \left( \log  \frac{M_t^2}{\mu^2}  - \frac{3}{2} \right)
+ 6 \, M_W^4 \left( \log  \frac{M_W^2}{\mu^2}  - \frac{5}{6} \right) \nonumber \\
&+& 3 \, {M}_Z^4 \left( \log \frac{{M}_Z^2}{\mu^2} - \frac{5}{6} \right) 
+M_h^4 \left( \log  \frac{M_h^2}{\mu^2}  - \frac{3}{2} \right)
+ 3 \, M_{\chi}^4 \left( \log  \frac{M_{\chi}^2}{\mu^2} - \frac{3}{2} \right) \n \\
& + & 2 \left(1 - \frac{1}{\xi_W} \right)^2 \, \left( \log \frac{M_W^2}{\mu^2}   \right)  \partial^4
+ 4 \left( 1 - \frac{1}{\xi_W} \right) \, \left( \log  \frac{M_W^2}{\mu^2}  - 1 \right) \, \partial^2 M_W^2 
\nonumber \\
& + &  \left(1 - \frac{1}{\xi_B} \right)^2 \, \left( \log \frac{M_Z^2}{\mu^2}  \right)  \partial^4
+ 2 \left( 1 - \frac{1}{\xi_B} \right) \, \left( \log  \frac{M_Z^2}{\mu^2}  - 1 \right) \, \partial^2 M_Z^2
\Bigg] \nonumber\,,
\end{eqnarray}
where 
\begin{eqnarray}
M_h^2 &=& -m^2 + 3 \lambda \phi^2 \,, 
~M_{\chi}^2 = -m^2 + \lambda \phi^2  \,, \nonumber\\
M_W^2 &=& \tfrac{1}{4} g^2 \phi^2 \,, ~
M_B^2 = \tfrac{1}{4} g'^{2} \phi^2 \,,~M_t^2 = \frac{y_t^2}{2} \phi^2  \,,
\end{eqnarray}
where $M_W^2=M_{W^i}^2 ,\;\; M_\chi^2=M_{\chi_i}^2$ $\forall \; i\in\{1,2,3\}$.
In this case, the gauge parameters $\xi_{W,B}$ are also associated with the total derivative terms, e.g., $\partial^4, \partial^2 M_W^2,  \partial^2 M_Z^2$. Thus, with a constant background, these terms do not contribute to the effective potential, which reads
\begin{eqnarray}\label{eq:SMVeff-HK-final} 
 V^{(1)}_{\rm{eff}}|^{\rm{HK}}_{\rm SM} &=& \frac{1}{64 \pi^2} \text{tr} \Bigg[ 
- M_t^4 \left( \log  \frac{M_t^2}{\mu^2} - \frac{3}{2} \right)
+ M_W^4 \left( \log \frac{M_W^2}{\mu^2} - \frac{5}{6} \right)  \\
&+& {M}_Z^4 \left( \log \frac{{M}_Z^2}{\mu^2}   - \frac{5}{6} \right) 
+M_h^4 \left( \log  \frac{M_h^2}{\mu^2}  - \frac{3}{2} \right)
+M_{\chi}^4 \left( \log  \frac{M_{\chi}^2}{\mu^2} - \frac{3}{2} \right) \Bigg] \nonumber \\
&=&  \frac{1}{64 \pi^2} \Bigg[ 
- 12 \, M_t^4 \left( \log  \frac{M_t^2}{\mu^2}  - \frac{3}{2} \right)
+ 6 \, M_W^4 \left( \log  \frac{M_W^2}{\mu^2}  - \frac{5}{6} \right) \nonumber \\
&+& 3 \, {M}_Z^4 \left( \log \frac{{M}_Z^2}{\mu^2} - \frac{5}{6} \right) 
+M_h^4 \left( \log  \frac{M_h^2}{\mu^2}  - \frac{3}{2} \right)
+ 3 \, M_{\chi}^4 \left( \log  \frac{M_{\chi}^2}{\mu^2} - \frac{3}{2} \right) 
\Bigg] \nonumber\,.
\end{eqnarray}
This gauge-independent potential matches the effective potential computed in the Landau gauge using the functional method, as shown in Eq.~\ref{eq:SMVeff-LG}.

\bibliographystyle{JHEP}
\bibliography{references}

\providecommand{\href}[2]{#2}\begingroup\raggedright\begin{thebibliography}{10}

\bibitem{Coleman:1973jx}
S.~R. Coleman and E.~J. Weinberg, {\it {Radiative Corrections as the Origin of
  Spontaneous Symmetry Breaking}},  {\em Phys. Rev. D} {\bf 7} (1973)
  1888--1910.

\bibitem{Sher:1988mj}
M.~Sher, {\it {Electroweak Higgs Potentials and Vacuum Stability}},  {\em Phys.
  Rept.} {\bf 179} (1989) 273--418.

\bibitem{Jackiw:1974cv}
R.~Jackiw, {\it {Functional evaluation of the effective potential}},  {\em
  Phys. Rev. D} {\bf 9} (1974) 1686.

\bibitem{Dolan:1974gu}
L.~Dolan and R.~Jackiw, {\it {Gauge Invariant Signal for Gauge Symmetry
  Breaking}},  {\em Phys. Rev. D} {\bf 9} (1974) 2904.

\bibitem{Kluberg-Stern:1974iel}
H.~Kluberg-Stern and J.~B. Zuber, {\it {Ward Identities and Some Clues to the
  Renormalization of Gauge Invariant Operators}},  {\em Phys. Rev. D} {\bf 12}
  (1975) 467--481.

\bibitem{Piguet:1984js}
O.~Piguet and K.~Sibold, {\it {Gauge Independence in Ordinary {Yang-Mills}
  Theories}},  {\em Nucl. Phys. B} {\bf 253} (1985) 517--540.

\bibitem{Kluberg-Stern:1974nmx}
H.~Kluberg-Stern and J.~B. Zuber, {\it {Renormalization of Nonabelian Gauge
  Theories in a Background Field Gauge. 1. Green Functions}},  {\em Phys. Rev.
  D} {\bf 12} (1975) 482--488.

\bibitem{Nielsen:1975fs}
N.~K. Nielsen, {\it {On the Gauge Dependence of Spontaneous Symmetry Breaking
  in Gauge Theories}},  {\em Nucl. Phys. B} {\bf 101} (1975) 173--188.

\bibitem{DelCima:1999gg}
O.~M. Del~Cima, D.~H.~T. Franco, and O.~Piguet, {\it {Gauge independence of the
  effective potential revisited}},  {\em Nucl. Phys. B} {\bf 551} (1999)
  813--825, [\href{http://arxiv.org/abs/hep-th/9902084}{{\tt hep-th/9902084}}].

\bibitem{Gambino:1999ai}
P.~Gambino and P.~A. Grassi, {\it {The Nielsen identities of the SM and the
  definition of mass}},  {\em Phys. Rev. D} {\bf 62} (2000) 076002,
  [\href{http://arxiv.org/abs/hep-ph/9907254}{{\tt hep-ph/9907254}}].

\bibitem{Grassi:2001bz}
P.~A. Grassi, B.~A. Kniehl, and A.~Sirlin, {\it {Width and partial widths of
  unstable particles in the light of the Nielsen identities}},  {\em Phys. Rev.
  D} {\bf 65} (2002) 085001, [\href{http://arxiv.org/abs/hep-ph/0109228}{{\tt
  hep-ph/0109228}}].

\bibitem{Goria:2011wa}
S.~Goria, G.~Passarino, and D.~Rosco, {\it {The Higgs Boson Lineshape}},  {\em
  Nucl. Phys. B} {\bf 864} (2012) 530--579,
  [\href{http://arxiv.org/abs/1112.5517}{{\tt arXiv:1112.5517}}].

\bibitem{Nielsen:2014spa}
N.~K. Nielsen, {\it {Removing the gauge parameter dependence of the effective
  potential by a field redefinition}},  {\em Phys. Rev. D} {\bf 90} (2014),
  no.~3 036008, [\href{http://arxiv.org/abs/1406.0788}{{\tt arXiv:1406.0788}}].

\bibitem{Aitchison:1983ns}
I.~J.~R. Aitchison and C.~M. Fraser, {\it {Gauge Invariance and the Effective
  Potential}},  {\em Annals Phys.} {\bf 156} (1984) 1.

\bibitem{Metaxas:1995ab}
D.~Metaxas and E.~J. Weinberg, {\it {Gauge independence of the bubble
  nucleation rate in theories with radiative symmetry breaking}},  {\em Phys.
  Rev. D} {\bf 53} (1996) 836--843,
  [\href{http://arxiv.org/abs/hep-ph/9507381}{{\tt hep-ph/9507381}}].

\bibitem{Patel:2011th}
H.~H. Patel and M.~J. Ramsey-Musolf, {\it {Baryon Washout, Electroweak Phase
  Transition, and Perturbation Theory}},  {\em JHEP} {\bf 07} (2011) 029,
  [\href{http://arxiv.org/abs/1101.4665}{{\tt arXiv:1101.4665}}].

\bibitem{Andreassen:2014eha}
A.~Andreassen, W.~Frost, and M.~D. Schwartz, {\it {Consistent Use of Effective
  Potentials}},  {\em Phys. Rev. D} {\bf 91} (2015), no.~1 016009,
  [\href{http://arxiv.org/abs/1408.0287}{{\tt arXiv:1408.0287}}].

\bibitem{Alexander:2008hd}
L.~P. Alexander and A.~Pilaftsis, {\it {The One-Loop Effective Potential in
  Non-Linear Gauges}},  {\em J. Phys. G} {\bf 36} (2009) 045006,
  [\href{http://arxiv.org/abs/0809.1580}{{\tt arXiv:0809.1580}}].

\bibitem{Garny:2012cg}
M.~Garny and T.~Konstandin, {\it {On the gauge dependence of vacuum transitions
  at finite temperature}},  {\em JHEP} {\bf 07} (2012) 189,
  [\href{http://arxiv.org/abs/1205.3392}{{\tt arXiv:1205.3392}}].

\bibitem{Espinosa:2015qea}
J.~R. Espinosa, G.~F. Giudice, E.~Morgante, A.~Riotto, L.~Senatore, A.~Strumia,
  and N.~Tetradis, {\it {The cosmological Higgstory of the vacuum
  instability}},  {\em JHEP} {\bf 09} (2015) 174,
  [\href{http://arxiv.org/abs/1505.04825}{{\tt arXiv:1505.04825}}].

\bibitem{Quiros:1999jp}
M.~Quiros, {\it {Finite temperature field theory and phase transitions}},  in
  {\em {ICTP Summer School in High-Energy Physics and Cosmology}},
  pp.~187--259, 1, 1999.
\newblock \href{http://arxiv.org/abs/hep-ph/9901312}{{\tt hep-ph/9901312}}.

\bibitem{Hirvonen:2021zej}
J.~Hirvonen, J.~L{\"o}fgren, M.~J. Ramsey-Musolf, P.~Schicho, and T.~V.~I.
  Tenkanen, {\it {Computing the gauge-invariant bubble nucleation rate in
  finite temperature effective field theory}},  {\em JHEP} {\bf 07} (2022) 135,
  [\href{http://arxiv.org/abs/2112.08912}{{\tt arXiv:2112.08912}}].

\bibitem{Lofgren:2021ogg}
J.~L{\"o}fgren, M.~J. Ramsey-Musolf, P.~Schicho, and T.~V.~I. Tenkanen, {\it
  {Nucleation at Finite Temperature: A Gauge-Invariant Perturbative
  Framework}},  {\em Phys. Rev. Lett.} {\bf 130} (2023), no.~25 251801,
  [\href{http://arxiv.org/abs/2112.05472}{{\tt arXiv:2112.05472}}].

\bibitem{Bernardo:2025vkz}
F.~Bernardo, P.~Klose, P.~Schicho, and T.~V.~I. Tenkanen, {\it
  {Higher-dimensional operators at finite-temperature affect gravitational-wave
  predictions}},  \href{http://arxiv.org/abs/2503.18904}{{\tt
  arXiv:2503.18904}}.

\bibitem{Elias-Miro:2014pca}
J.~Elias-Miro, J.~R. Espinosa, and T.~Konstandin, {\it {Taming Infrared
  Divergences in the Effective Potential}},  {\em JHEP} {\bf 08} (2014) 034,
  [\href{http://arxiv.org/abs/1406.2652}{{\tt arXiv:1406.2652}}].

\bibitem{Espinosa:2016uaw}
J.~R. Espinosa, M.~Garny, and T.~Konstandin, {\it {Interplay of Infrared
  Divergences and Gauge-Dependence of the Effective Potential}},  {\em Phys.
  Rev. D} {\bf 94} (2016), no.~5 055026,
  [\href{http://arxiv.org/abs/1607.08432}{{\tt arXiv:1607.08432}}].

\bibitem{Balui:2025kat}
D.~Balui, J.~Chakrabortty, D.~Dey, and S.~Mohanty, {\it {Gauge invariant
  effective potential}},  {\em Phys. Rev. D} {\bf 111} (2025), no.~8 085032,
  [\href{http://arxiv.org/abs/2502.17156}{{\tt arXiv:2502.17156}}].

\bibitem{Hawking:1976ja}
S.~W. Hawking, {\it {Zeta Function Regularization of Path Integrals in Curved
  Space-Time}},  {\em Commun. Math. Phys.} {\bf 55} (1977) 133.

\bibitem{Elizalde2}
E.~Elizalde, {\it Zeta-function regularization techniques for series summations
  and applications},  in {\em Proceedings Leipzig Workshop Quantum field theory
  under the influence of external conditions}, (Leipzig), Universit{\"a}t
  Leipzig, Naturwissenschaftlich-Theoretisches Zentrum, 10, 1992.

\bibitem{Elizalde-3}
E.~Elizalde, {\it {Complete determination of the singularity structure of zeta
  functions}},  {\em J. Phys. A} {\bf 30} (1997) 2735--2744,
  [\href{http://arxiv.org/abs/hep-th/9608056}{{\tt hep-th/9608056}}].

\bibitem{Wodzicki}
M.~Wodzicki, {\em Noncommutative residue Chapter I. Fundamentals},
  pp.~320--399.
\newblock Springer Berlin Heidelberg, Berlin, Heidelberg, 1987.

\bibitem{Mickelsson:1994fb}
J.~Mickelsson, {\it {Wodzicki residue and anomalies of current algebras}},
  {\em Lect. Notes Phys.} {\bf 436} (1994) 123--135,
  [\href{http://arxiv.org/abs/hep-th/9404093}{{\tt hep-th/9404093}}].

\bibitem{Bytsenko:1994bc}
A.~A. Bytsenko, G.~Cognola, L.~Vanzo, and S.~Zerbini, {\it {Quantum fields and
  extended objects in space-times with constant curvature spatial section}},
  {\em Phys. Rept.} {\bf 266} (1996) 1--126,
  [\href{http://arxiv.org/abs/hep-th/9505061}{{\tt hep-th/9505061}}].

\bibitem{Elizalde:1997nd}
E.~Elizalde, L.~Vanzo, and S.~Zerbini, {\it {Zeta function regularization, the
  multiplicative anomaly and the Wodzicki residue}},  {\em Commun. Math. Phys.}
  {\bf 194} (1998) 613--630, [\href{http://arxiv.org/abs/hep-th/9701060}{{\tt
  hep-th/9701060}}].

\bibitem{Elizalde:1998xq}
E.~Elizalde, A.~Filippi, L.~Vanzo, and S.~Zerbini, {\it {Is the multiplicative
  anomaly dependent on the regularization?}},
  \href{http://arxiv.org/abs/hep-th/9804071}{{\tt hep-th/9804071}}.

\bibitem{Seeley}
R.~T. Seeley, {\it {Complex powers of an elliptic operator}},  {\em Proc. Symp.
  Pure Math.} {\bf 10} (1967) 288--307.

\bibitem{Belkov:1995gjw}
A.~A. Bel'kov, A.~V. Lanyov, and A.~Schaale, {\it {Calculation of heat kernel
  coefficients and usage of computer algebra}},  {\em Comput. Phys. Commun.}
  {\bf 95} (1996) 123--130, [\href{http://arxiv.org/abs/hep-ph/9506237}{{\tt
  hep-ph/9506237}}].

\bibitem{Vassilevich:2003xt}
D.~V. Vassilevich, {\it {Heat kernel expansion: User's manual}},  {\em Phys.
  Rept.} {\bf 388} (2003) 279--360,
  [\href{http://arxiv.org/abs/hep-th/0306138}{{\tt hep-th/0306138}}].

\bibitem{Avramidi:2001ns}
I.~G. Avramidi, {\it Heat kernel approach in quantum field theory},  in {\em
  Nucl. Phys. B Proc. Suppl.}, vol.~104, pp.~3--32, 2002.
\newblock \href{http://arxiv.org/abs/math-ph/0107018}{{\tt math-ph/0107018}}.

\bibitem{Kontsevich:1994xe}
M.~Kontsevich and S.~Vishik, {\it {Geometry of determinants of elliptic
  operators}},  \href{http://arxiv.org/abs/hep-th/9406140}{{\tt
  hep-th/9406140}}.

\bibitem{Avramidi:1990je}
I.~G. Avramidi, {\it {The Covariant Technique for Calculation of One Loop
  Effective Action}},  {\em Nucl. Phys. B} {\bf 355} (1991) 712--754. [Erratum:
  Nucl.Phys.B 509, 557--558 (1998)].

\bibitem{Banerjee:2023iiv}
U.~Banerjee, J.~Chakrabortty, S.~U. Rahaman, and K.~Ramkumar, {\it {One-loop
  effective action up to dimension eight: integrating out heavy scalar(s)}},
  {\em Eur. Phys. J. Plus} {\bf 139} (2024), no.~2 159,
  [\href{http://arxiv.org/abs/2306.09103}{{\tt arXiv:2306.09103}}].

\bibitem{Chakrabortty:2023yke}
J.~Chakrabortty, S.~U. Rahaman, and K.~Ramkumar, {\it {One-loop effective
  action up to dimension eight: Integrating out heavy fermion(s)}},  {\em Nucl.
  Phys. B} {\bf 1000} (2024) 116488,
  [\href{http://arxiv.org/abs/2308.03849}{{\tt arXiv:2308.03849}}].

\bibitem{Banerjee:2023xak}
U.~Banerjee, J.~Chakrabortty, S.~U. Rahaman, and K.~Ramkumar, {\it {One-loop
  effective action up to any mass-dimension for non-degenerate scalars and
  fermions including light{\textendash}heavy mixing}},  {\em Eur. Phys. J.
  Plus} {\bf 139} (2024), no.~2 169,
  [\href{http://arxiv.org/abs/2311.12757}{{\tt arXiv:2311.12757}}].

\bibitem{Banerjee:2024rbc}
U.~Banerjee, J.~Chakrabortty, and K.~Ramkumar, {\it {Renormalization of scalar
  and fermion interacting field theory for arbitrary loop:
  Heat{\textendash}Kernel approach}},  {\em Eur. Phys. J. Plus} {\bf 139}
  (2024), no.~8 714, [\href{http://arxiv.org/abs/2404.02734}{{\tt
  arXiv:2404.02734}}].

\bibitem{Adhikary:2025pbb}
N.~Adhikary, J.~Das, and D.~Dey, {\it {Two-loop dimension Six Effective Action:
  Integrating Out Heavy Scalar}},  \href{http://arxiv.org/abs/2501.01313}{{\tt
  arXiv:2501.01313}}.

\bibitem{Adhikary:2025gdh}
N.~Adhikary, T.~Biswas, J.~Chakrabortty, C.~Englert, and M.~Spannowsky, {\it
  {Electroweak Scalar Effects Beyond Dimension-6 in SMEFT}},
  \href{http://arxiv.org/abs/2501.12160}{{\tt arXiv:2501.12160}}.

\bibitem{Urbano:2019ohp}
A.~Urbano, {\it {Inflation without gauge redundancy}},  {\em JCAP} {\bf 04}
  (2020) 040, [\href{http://arxiv.org/abs/2001.05480}{{\tt arXiv:2001.05480}}].

\bibitem{Fukuda:1975di}
R.~Fukuda and T.~Kugo, {\it {Gauge Invariance in the Effective Action and
  Potential}},  {\em Phys. Rev. D} {\bf 13} (1976) 3469.

\bibitem{Andreassen:2014gha}
A.~Andreassen, W.~Frost, and M.~D. Schwartz, {\it {Consistent Use of the
  Standard Model Effective Potential}},  {\em Phys. Rev. Lett.} {\bf 113}
  (2014), no.~24 241801, [\href{http://arxiv.org/abs/1408.0292}{{\tt
  arXiv:1408.0292}}].

\bibitem{Espinosa:2016nld}
J.~R. Espinosa, M.~Garny, T.~Konstandin, and A.~Riotto, {\it {Gauge-Independent
  Scales Related to the Standard Model Vacuum Instability}},  {\em Phys. Rev.
  D} {\bf 95} (2017), no.~5 056004,
  [\href{http://arxiv.org/abs/1608.06765}{{\tt arXiv:1608.06765}}].

\bibitem{Martin:2014bca}
S.~P. Martin, {\it {Taming the Goldstone contributions to the effective
  potential}},  {\em Phys. Rev. D} {\bf 90} (2014), no.~1 016013,
  [\href{http://arxiv.org/abs/1406.2355}{{\tt arXiv:1406.2355}}].

\bibitem{Barvinsky:2024irk}
A.~O. Barvinsky and A.~E. Kalugin, {\it {Notes on peculiarities of the
  Schwinger-DeWitt technique: One-loop double poles, total-derivative terms,
  and determinant anomalies}},  {\em Phys. Rev. D} {\bf 110} (2024), no.~10
  105007, [\href{http://arxiv.org/abs/2408.16174}{{\tt arXiv:2408.16174}}].

\bibitem{Bytsenko:2003}
A.~A. Bytsenko, G.~Cognola, E.~Elizalde, V.~Moretti, and S.~Zerbini, {\em
  Analytic Aspects of Quantum Fields}.
\newblock World Scientific, Singapore, 2003.

\bibitem{Seeley:1967ea}
R.~T. Seeley, {\it {Complex powers of an elliptic operator}},  {\em Proc. Symp.
  Pure Math.} {\bf 10} (1967) 288--307.

\bibitem{Dolan:1973qd}
L.~Dolan and R.~Jackiw, {\it {Symmetry Behavior at Finite Temperature}},  {\em
  Phys. Rev. D} {\bf 9} (1974) 3320--3341.

\bibitem{Megias:2003ui}
E.~Megias, E.~Ruiz~Arriola, and L.~L. Salcedo, {\it {The Thermal heat kernel
  expansion and the one loop effective action of QCD at finite temperature}},
  {\em Phys. Rev. D} {\bf 69} (2004) 116003,
  [\href{http://arxiv.org/abs/hep-ph/0312133}{{\tt hep-ph/0312133}}].

\bibitem{Chakrabortty:2024wto}
J.~Chakrabortty and S.~Mohanty, {\it {One Loop Thermal Effective Action}},
  \href{http://arxiv.org/abs/2411.14146}{{\tt arXiv:2411.14146}}.

\bibitem{Moral-Gamez:2011wcb}
F.~J. Moral-Gamez and L.~L. Salcedo, {\it {Derivative expansion of the heat
  kernel at finite temperature}},  {\em Phys. Rev. D} {\bf 85} (2012) 045019,
  [\href{http://arxiv.org/abs/1110.6300}{{\tt arXiv:1110.6300}}].

\bibitem{DiLuzio:2014bua}
L.~Di~Luzio and L.~Mihaila, {\it {On the gauge dependence of the Standard Model
  vacuum instability scale}},  {\em JHEP} {\bf 06} (2014) 079,
  [\href{http://arxiv.org/abs/1404.7450}{{\tt arXiv:1404.7450}}].

\end{thebibliography}\endgroup
\end{document}